\newif\iflong\longfalse
 \documentclass[12pt]{article} 
 \usepackage{fullpage}
\usepackage{amsmath,amsfonts,amsthm, amssymb,bm}
\usepackage{mathtools}
\usepackage{prftree}
\usepackage{xcolor}
\usepackage{xspace}
\usepackage{nameref}
\usepackage{hyperref}
\usepackage{xifthen}
\usepackage{mathpartir}
\usepackage{cancel}

\definecolor{green}{rgb}{0,0,0}
\definecolor{red}{rgb}{0,0,0}
\definecolor{pink}{rgb}{0,0,0}
\definecolor{rred}{rgb}{0,0,0}
\definecolor{orange}{rgb}{0,0,0}

\newtheorem{thm}{Theorem}
\newtheorem{lm}[thm]{Lemma}

\theoremstyle{definition}
\newtheorem{df}[thm]{Definition}

\newtheorem{rem}[thm]{Remark}
\newtheorem{ex}[thm]{Example}

\newcommand{\new}{\bm{\nu}}
\newcommand{\trans}[1]{\xrightarrow{#1}}
\newcommand{\sep}{\ \big| {\ }}
\newcommand{\vfopp}[1][]{\textcolor{red}{\Gamma#1}} 
\newcommand{\vfpl}[1][]{\textcolor{green}{\Delta#1}} 
\newcommand{\Vset}{\ensuremath{A}} 

\newcommand{\ser}[1]{#1 \,\rotatebox[origin=c]{270}{$\rightsquigarrow$}}
\newcommand{\stack}[1][]{\textcolor{green}{{\sigma}{#1}}} 
\newcommand{\kcats}[1][]{\textcolor{red}{{\kappa}{#1}}} 
\newcommand{\resmin}[2]{#1^{-#2}}
\newcommand{\arr}{\mapsto} 
\newcommand{\cur}{\textcolor{pink}{\ast}} 
\newcommand{\cons}{;} 
\newcommand{\ite}[3]{\mathtt{if}\ #1\ \mathtt{then}\ #2 \ \mathtt{else}\ #3}
\newcommand{\out}[3]{\overline{#1}\ifthenelse{\equal{#2}{}}{}{\langle #2 \rangle} \ifthenelse{\equal{#3}{}}{}{(#3)}}
\newcommand{\inp}[2]{#1(#2)}
\newcommand{\inpA}[3]{#1\ifthenelse{\equal{#2}{}}{}{\langle #2 \rangle} \ifthenelse{\equal{#3}{}}{}{(#3)}}

\newcommand{\typpl}[2]{
 \textcolor{green}{#1}
}
\newcommand{\typopp}[2]{\langle \textcolor{red}{#1} | \textcolor{red}{#2} \rangle}
\renewcommand{\typopp}[2]{ \textcolor{red}{#1}  }
\newcommand{\typ}[2]{#1\vdash #2}
\newcommand{\bisim}[3]{#1 #2 \approx #3}
\renewcommand{\bisim}[3]{ #2 \mathrel{\approx^{#1}} #3}
\newcommand{\typtrans}[5]{\typPAIR{#1}{#2} \xrightarrow{#3} \typPAIR{#4}{#5}}
\renewcommand{\typtrans}[5]{\typPAIR{#1}{#2} \trans{#3}\typPAIR{#4}{#5}}
\newcommand{\typTrans}[5]{\typPAIR{#1}{#2} \xRightarrow{\hat{#3}}\typPAIR{#4}{#5}}

\newcommand{\typTRANS}[5]{\typPAIR{#1}{#2} \xRightarrow{{#3}}\typPAIR{#4}{#5}}

\newcommand{\typRed}[4]{\typPAIR{#1}{#2} \Longrightarrow \typPAIR{#3}{#4}}
\newcommand{\typPAIR}[2]{[#1]\, #2}

\renewcommand{\typTrans}[5]{[#1 ] #2 \xRightarrow{\hat{#3}} [#4 ] #5}

\newcommand{\underoppD}[3]{(#1 , #2,#3)}
\newcommand{\co}[1]{#1^\texttt{O}}
\newcommand{\ci}[1]{#1^\texttt{I}}

\newcommand{\bfcdotB}{ {\mbox{\boldmath $.$}}  }

\newcommand{\barbedbis}
{\mathrel{\stackrel{\bfcdotB}{\approx}}}

\newcommand{\conteq}[2]{\mathrel{\cong}_{#1}^{#2}}
\newcommand{\conteqNOtype}{\mathrel{\cong}}

\newcommand{\bomega}{\ensuremath{\downarrow_{\omega}}}
\newcommand{\nbomega}{\ensuremath{\not\,\bomega}}
\newcommand{\dom}[1]{\mathtt{dom}(#1)}
\newcommand{\fnames}[1]{\ensuremath{\mathtt{fn}(#1)}}
\newcommand{\fNho}[1]{\ensuremath{\mathtt{nHO}(#1)}}
\newcommand{\bnames}[1]{\ensuremath{\mathtt{bn}(#1)}}

\newcommand{\Bsplit}[1]{\mathtt{split}(#1)}

\newcommand{\rname}[1]{\ensuremath{\mathtt{#1}}}
\newcommand{\justif}{\ensuremath{ \curvearrowleft}}
\newcommand{\njustif}{\ensuremath{\not\!\curvearrowleft}}
\newcommand{\view}[2]{\ensuremath{\mathtt{view}_{#1}(#2)}}
\newcommand{\trace}{\ensuremath{\mathcal T}}
\newcommand{\prune}[3]{\ensuremath{\mathtt{Pr}_{#1}^{#2}(#3)}}

\newcommand{\meets}[2]{\ensuremath{#1^\smallfrown#2}}
\newcommand{\sdash}{\mbox{-}}
\newcommand{\eval}[2]{\ensuremath{#1\Downarrow#2}}

\newcommand{\daniel}[1]{\textcolor{orange}{\textbf{DH:}#1}}
\newcommand{\iwan}[1]{\textcolor{orange}{\textbf{IQ:}#1}}
\newcommand{\TODO}[1]{\textcolor{rred}{\textbf{TODO:}#1}}

\def\nil{{\boldsymbol 0}} 
\def\res#1{{\boldsymbol \nu} #1\:}   

\DeclareUnicodeCharacter{03C0}{$\pi$}

\usepackage{davidemacros}

\renewcommand{\nilact}{\ensuremath{\nil_{\bm{\cur}}}}

\renewcommand{\finish}[1]{}
\renewcommand{\daniel}[1]{}
\renewcommand{\iwan}[1]{}
\renewcommand{\TODO}[1]{}

\title{
 First-order store and visibility in name-passing calculi
}
\date{}
\author{Daniel Hirschkoff \and Iwan Quémerais \and Davide Sangiorgi}

\begin{document}

\maketitle

\begin{abstract}
  The $\pi$-calculus is the paradigmatical name-passing calculus.
  While being purely name-passing, it allows the representation of
  higher-order functions and store.

We study how $\pi$-calculus processes can be
  controlled so that computations can only involve storage of
  first-order values. 
  The discipline 
  is enforced by a type system that is based on the notion of
  visibility, coming from game semantics.  We discuss the impact of
  visibility on the behavioural theory.  We propose
  characterisations of may-testing and barbed equivalence, based on
  (variants of) trace equivalence and labelled bisimilarity, in
  the case where computation is sequential, and in the case where
  computation is well-bracketed.
\end{abstract}

\section{Introduction}
\label{s:intro}

The $\pi$-calculus is one of the best known
 process calculi. It has a handful of operators~---
parallelism, input, output, restriction being the main ones~---
with which it achieves 
an amazing expressive power. Yet, it has  a  rich
algebraic theory and a wide spectrum of proof techniques.
In the $\pi$-calculus, 
computation is communication, by means of message-passing
synchronisations between processes. Names are the communication units;
new names may be created, and may themselves  be exchanged in
communications. 
When used for modelling, $\pi$-calculus is normally typed, so to be
able to support structured data values,  
beginning with  basic first-order values such as  integers and booleans. 

The $\pi$-calculus has been advocated as a model to interpret, and
give semantics to, languages with higher-order features.  
For instance, the representation of functions as $\pi$-calculus
processes has been widely studied, beginning with Milner's seminal work
\cite{MilnerFunsProcs}. 
Usually, higher-order
languages include forms of \emph{store}.
It is therefore important to understand how expected semantic
properties of store can be captured in a language for pure concurrency 
such as the $\pi$-calculus.

An aspect of store with  strong semantic consequences is the
distinction between higher-order and first-order store. 
The term `first-order store' refers to a memory
model where variables can hold only basic data types such as integers
and  booleans. This contrasts with a higher-order store, where
variables can also  store functions, or code, or references to these.
Limiting the store to first-order data makes it easier to reason
about program behaviours, and verify behavioural properties.  It may
also enhance security and predictability guarantees by preventing
unintended behaviours associated with dynamic function storage and code
injection.  A striking example of a property that may be broken with
higher-order store is termination.  For instance, the simply-typed
$\lambda$-calculus is strongly normalising and therefore terminating.
This property is maintained by the addition of first-order store, but
not with higher-order store, as functions may recursively reference
and invoke each other through the store.
In game semantics of higher-order languages, the distinction between
first-order and higher-order store has led to studying the concept of
\emph{visibility}.  Intuitively, visibility ensures us that each move
in a computational game is justified by prior interactions.  With the
introduction of higher-order store, the  visibility condition has to 
be relaxed, reflecting the added complexity in tracking information
flow within a program.

The goal of this paper is precisely to understand the meaning of
first-order store and visibility within the $\pi$-calculus. 
In functional languages based on the $\lambda$-calculus, imposing
first-order store is simply determined by the constraint that all
references may only store first-order values. 
In the $\pi$-calculus, in contrast, there is no explicit notion of
reference,  and therefore the meaning of `first-order store' requires
some care. 
In $\pi$-calculi, a piece of store is usually represented by an output
message. For instance $\outDS h 5$ may be thought of as the statement that
name $h$  currently contains $5$. A process that inputs at $h$ and
then re-emits a message at $h$, such as $\inp h z. \outDS h {z+1}$, is a
process that acts on such a  store. 
In this case, $h$ is a first-order name, hence $h$ implements a piece
of first-order store. 
An example of higher-order store is 
given by the process 
\begin{equation}
\label{e:P}
P \defi \inp a {b} . ( ( \boutDS b d . ! d. R )  \,|\,  c . \boutDS b
{d'}. d'.R'  )
\end{equation}
 Here,
  $\boutDS b d$ is a  bound output, and represents the output of a
 `new' name $d$, and similarly for 
   $\boutDS b {d'}$; whereas, e.g., 
$d.R$ and $c. Q$ represent inputs in
which no value is exchanged (i.e., $d$ and $c$  carry {\tt unit} values). 
In $P$, name $a$ is higher-order, in that $a$ carries a (pointer to a)
service. 
In the input at $a$, the service name received in the input (the
parameter $b$) is used after the input, but it is also 
 stored, and used later when
$c$ is invoked. 
In other words, the visibility of $c$ (the services that $c$ may call
when invoked) depends on interactions that have occurred earlier, at
$a$.  For this reason, $P$ implements a form of higher-order store. 
When this does not happen, that is, 
the visibility of a higher-order name (the set of services that 
may be called when the name is invoked)
 may be determined at the point
in which the name is created, the process implements  a form of  first-order
store. (The word `service' is freely used, here and elsewhere, to indicate
the input end of a name; in $\pi$-calculus,  a name may have
multiple input occurrences, in arbitrary syntactic positions within a
process, in contrast with, say,  the  function declarations in 
functional languages, which are supposed to be    unique.)


The amazing expressive power of the $\pi$-calculus 
 also   makes 
the contexts of the calculus very discriminating;  
as a consequence,  
behavioural equivalences,
which are supposed to be preserved by
the contexts of the calculus,  
 are rather demanding relations.
A  well-established way of
  imposing constraints on the set of legal contexts,
 thus 
obtaining  usefully coarser 
behavioural equivalences,
is to adopt behavioural type systems.
Types may be used   in order 
to capture  
 communication patterns  on the usage of names, such as  
  capabilities, linearity,  sessions, and so on
on, e.g.,~\cite{DBLP:journals/mscs/PierceS96,DBLP:journals/toplas/KobayashiPT99,DBLP:conf/esop/HondaVK98,AnconaBB0CDGGGH16}. 
Type   systems have also been
designed to capture specific properties of processes, such as
termination, deadlock-freedom, lock-freedom, e.g.,
\cite{DBLP:journals/iandc/YoshidaBH04,
  DBLP:journals/toplas/KobayashiS10, DBLP:conf/csl/Padovani14}.

A further  step is then  to tune the proof techniques of the $\pi$-calculus
to such type systems, so to be able to actually prove the behavioural equalities
that only hold in presence of types. 
For instance, in the case of may testing (or contextual equivalence)
this may yield to  refinements or modifications of the notion of trace equivalence; 
whereas 
in the case of barbed equivalence this may lead to 
 refinements or modifications  of    the standard
notion of labelled bisimilarity.
The resulting proof techniques 
must be sound with respect to
initial contextually-defined relations; 
ideally, they should also be complete.

In this paper we propose a type system that guarantees  first-order
store in $\pi$-calculus processes. The type system makes use of
\emph{visibility functions}, that is, partial finite functions that
specify, for each higher-order name $b$ that may occur free in a process,
the set of services (i.e.,  higher-order names)
 that may be invoked when a call at
$b$ is made. 
In order to better understand  the behavioural effects of
 visibility,  we combine it first  with \emph{sequentiality} and
then with \emph{well-bracketing}. 
`Sequentiality' intuitively indicates the existence of a 
 single thread of computation. 
That is, at any point there is a single
 process that is active and
  decides what the next computation step can be.
In the  encodings of the $\lambda$-calculus ~\cite{DBLP:journals/mscs/Milner92,DBLP:journals/iandc/Sangiorgi94}, 
a process is active, i.e., it carries the thread,  when it contains an
unguarded output at a higher-order name. 
When encoding $\lambda$-calculi with first-order references, an active
process may also expose the thread by exhibiting an input at a
first-order name. 
We  follow such a convention also in the paper: higher-order names
carry the thread via outputs, first-order names via inputs (other
choices technically would however   be possible). 
Sequentiality allows us to expose, in isolation,
the causality  properties of service calls, within a standard
(interleaving) semantics. 
These aspects would be  partially or totally obscured by the
 presence of other concurrent threads.

In higher-order languages, the effects of having first-order store are
particularly strong in absence of 
control operators; that is, 
using a  terminology
 borrowed from
 game semantics, in a 
'well-bracketed' setting, 
where the call-return interaction behaviour
 between a term and its context follows a stack discipline.
We therefore also consider the refinement of our visibility-based type
system under the well-bracketing hypothesis. 
To see an example of the effect of visibility, consider
the following well-bracketed example: 
\[ 
\res h (
\,
 \outDS h 0 ~|~ 
  \boutDS  a {c,q} . (
 ! c. h(z). \outDS h {z+1} |  
q(y). Q ) \,)
\]
Here, an external service $a$ is called, exporting a
 local service $c$ capable of  incrementing a private reference $h$,
 and with a return channel $q$. 
If visibility holds, then 
 the external environment cannot 
 store $c$. Therefore,  after providing  an answer at $q$ ,
the environment 
is unable to 
 call $c$ again,
thus incrementing $h$, unless an access to $c$ is explicitly provided
again (by process $Q$). 
This property may be used to perform optimisations, possibly including
 garbage-collection of the service $c$ if at some point $c$
is not used anymore in $Q$.

We then study characterisations for both may testing and barbed
equivalence, as special forms of trace equivalence and of labelled
bisimilarity. Following the tradition of testing, the 
language includes special success signals, that are used by  testers
 (i.e, processes running in parallel with the testees), 
and may be used at any time, to report  success.

When developing such characterisations, some care is necessary so to
distinguish between the visibility of the tested and of the tester
processes, as they need not be the same. 
For this,  we introduce  visibility-constrained LTSs, 
in which process transitions are constrained by a visibility
function, 
intuitively expressing the possibility for processes to perform
a certain  transition given 
 a certain
visibility function for the external observer.
As a consequence of
the transition,
the visibility function may need to be updated, 
so to yield  the visibility function  for the
next transition. 
In the case of well-bracketing, the constraints given by visibility
have to be coupled with those given by the stack names used to track
well-bracketing 
(and again, a distinction has to be made between the stacks for
the testers and those for the testees).
Based on these LTSs, we introduce forms of 
 trace equivalence and of labelled bisimilarity that 
are proved to be sound and complete, for may testing and barbed
equivalence, both in the case of sequentiality and in that of
well-bracketing.


We present the version of the $\pi$-calculus with higher-order and
first-order names in Section~\ref{s:lang}. In Section~\ref{s:types},
we define the type system implementing visibility for sequential
processes. We define typed labelled transitions and present  labelled
characterisations of behavioural
equivalences in Section~\ref{s:labels}.
Section~\ref{s:wb} is devoted to the extension with well-bracketing.

\section{Background: the $\pi$-calculus language}
\label{s:lang}


The $\pi$-calculus  language that will be used is called, technically,   Internal $\pi$ (\piI)~\cite{internalpi} 
(all channels exchanged are fresh -- 
 no
free object outputs), with the addition of  \emph{first-order  values} (such as integers and booleans), and expressions
that evaluate  
to first-order values, called \emph{first-order expressions}. 
 We do not specify what exactly first-order values and first-order
 expressions are.  
 We simply assume that, in a closed process,  a first-order expression evaluates to
 a first-order value. 
Moreover, 
only the output capability of names may be exported.
This means 
that, for instance, in an input $ \inp a {b}.P$ 
 name $b$ may only be used in output in $P$.
This constraint is often followed in theory and applications of the
$\pi$-calculus, as it simplifies various technical matters.

We assume  that names are distinguished (partitioned) into
\emph{higher-order} names and \emph{first-order} names, and
\emph{success names}. 
Higher-order names are the ordinary \piI names;  first-order names may
only carry first-order values, and they may not be passed around
(the second constraint simplifies the technical details, though it is not
mandatory, see Section~\ref{s:types}). 
In later sections, first-order names will be used to represent
references. The fact that these names may only carry first-order
values, together with constraints on visibility, will   
ensure that the calculus does not have (implicit or explicit forms of) higher-order store. 
The \emph{success} names are used,  in the
tradition of testing equivalences,  by tester processes
so to report results of experiments on
tested processes (thus,  the processes compared 
in the examples of the paper do not contain success names).

The calculi in the paper will be typed. 
For simplicity 
we define our type systems as
refinements 
of the  most basic type system for $\pi$-calculus, namely 
 Milner's {\em sorting} \cite{Mil91}, in which 
names  are partitioned into a collection of {\em types} (or sorts), and  
a sorting  function maps types onto types. 
We  assume that there is a sorting system under which all
processes we manipulate are well-typed.  
Thus, the sorting system
separates higher-order and first-order names;
moreover  we assume that
first-order names are monadic, and that 
higher-order names carry a pair of a  first-order value and a
higher-order
 name. 
 %
Thus $\fbout  a v{b} . P$ is the  output  of the first-order  value  $v $ and of
the fresh  higher-order name $b$ at  the (higher-order) name $a$, with
continuation $P$.
When writing examples,
we allow different sortings; for instance, 
writing $a.P$ and  $\out b{}{}.Q$ for inputs and 
 outputs in which no value is exchanged.

\begin{table}
\[ 
\begin{array}{rrrr}
  \mboxD{Higher-order names, \Nho}&  a,b,c
  &
    \hspace{1cm}~
  \mboxD{First-order expressions}&  e,f
  \\
    \mboxD{First-order names, \Nfo}&  h,k
                         &
  \mboxD{First-order variables}&  x,y
                                    \\
  \mboxD{Success names}&  \omega
                                 &
\mboxD{First-order values}&  v,w  
\end{array}
\] \[\begin{array}{rrclrrcl}
  \mboxD{Inputs}&  \alpha  &::= & 
                                  \inp a {x,b}\midd \inp h x
                                  \hspace{.51cm}~
                                &
\mboxD{Outputs}&  \outC \alpha  &::= & 
\fbout a e{b} \midd
\fout h e \midd \omega
 \\
  \mboxD{Processes}&  P,Q,R &::= &
\multicolumn{5}{l}{\alpha . P  \midd  \outC \alpha . P  \midd 
! \alpha . P
\midd
  P|Q\midd 
 \res a P
                                   \midd \res h P}
       \\ & & &
                \multicolumn{5}{l}{\midd  \nil
       \midd P+Q
  \midd \ifte efPQ}
\end{array}
 \]
\caption{The syntax of the calculus}
\label{ta:syntax}
\end{table} 

The syntactic categories of the calculus are presented in
Table~\ref{ta:syntax}. 
As  usual,  input and restriction are binding constructs, and give
rise to the expected definitions of free and bound names. 
An expression or a process is \emph{closed} if it does not have free
(first-order) variables. 
It is intended that transitions are defined on closed
processes.  

	The definition of structural congruence, written $\equiv$,  
and of the strong and weak  (untyped) labelled
        transitions, written
        $\arrD{\mu}$, $\Longrightarrow$, and $\Arr{\hat\mu}$,  are standard and are given in
 Appendix~\ref{a:calc} (later we will introduce typed variants of the LTS). 
We note $\fnames{P}$ (resp. $\fnames{\mu}$) the set of free names of $P$ (resp. $\mu$).
We sometimes abbreviate reductions  $P \arrD\tau P' $ as $P \longrightarrow P'$.


As behavioural equivalences we  consider both  
 may-testing and barbed equivalence. 
Both relations are contextual, in that they are defined by requiring 
a  closure under all `observers'. They both use `barbs' (or 'success')
signals;  
barbed  equivalence, in addition,  
makes use of   a bisimulation game on reductions. 
We write 
$\Dwa_\omega$ to indicate   
the possibility   of exhibiting an  $\omega$ barb, possibly
 after some reductions; i.e.,  $P \Dwa_\omega$ holds if there is
 $P'$
 with $P  \Longrightarrow P' $ and $P' $ has an unguarded occurrence
 of $\omega$.
Thus intuitively two processes are behaviourally equal  if they
can produce `the same barbs'  under all legal observers running
in parallel with them.
As we are in a typed setting, observer and tested processes should
have compatible typings. 
The meaning of 'compatible'  depends on the specific type
system. 
For  visibility, the meaning will be specified in Sections~\ref{s:types}
and \ref{s:wb}. 
We write 
$\typedOK \Delta P$, and say that $P$ is a \emph{$\Delta $-process},   if 
$P$ is typable under 
$\Delta$, and 
$\typedOK \Delta{ P,Q}$  if  both 
$\typedOK \Delta P$ and $\typedOK \Delta Q$ hold.
Behavioural equivalences are defined on closed processes; they are
extended to open processes (i.e., processes with  free first-order
variables) in the usual manner,  by  considering  all possible
closing substitutions. 
We only consider \emph{weak} behavioural equivalences (i.e., equivalences that abstract
from internal moves of processes, the reductions), as these are, in practice, the most
useful ones.  

\begin{df}[May testing]
\label{d:mt}
Suppose that  $\Gamma, \Delta $ are
compatible. Then two $\Delta $-processes $P$ and $Q$
 are \emph{may testing equivalent at 
$ \Gamma$},
written $P \conteq \Delta  \Gamma Q$, if
 for all $R$-processes 
 and 
success
  name $\omega $, we have $(R | P) \Dwa_\omega$ iff
$(R | Q) \Dwa_\omega$.
\end{df}

\begin{df}[Barbed bisimilarity and equivalence]
\label{d:bb} {\em Barbed bisimilarity}
is the largest symmetric relation $\wbbd{}$ 
s.t.\ 
  $P \wbbd{}  Q$ implies: 
\begin{enumerate}
\item   whenever $P  \longrightarrow  P'$, 
  there exists $Q'$ such that
 $Q \Longrightarrow Q'$ and $P' \wbbd{}  Q'$;

 \item   for each   $\omega$, if 
 $ {P \Downarrow_\omega}$ then also 
 $ {Q \Downarrow_\omega}$.
\end{enumerate}
Suppose that $\vfopp, \vfpl$ are compatible. Then two
$\vfpl$-processes $P$ and $Q$ are {\em barbed equivalent at
  $\vfopp$}, written
$P \wbe \vfpl \Gamma    Q$, if   
for all $R$ with $\typedOK \vfopp { R}$ 
it holds that  $ P |R  \wbbd{}
Q |R
  $.
\end{df}
We write 
$P \conteqNOtype  Q$  
(resp.\ $P  \wbeNOtype    Q$) 
when 
$P \conteq \Delta  \Gamma Q$ 
(resp.\ $P \wbe \Delta \Gamma    Q$) 
 holds under any typing $ \Delta $ for $P,Q$  and any $ \Gamma
$ compatible with $ \Delta $.  



\finish{maybe define also full barbed congruence ? and full contextual
  equivalence?  explain that these are more complicated}







We now introduce some notations that will be useful below.
In the paper we use partial
functions whose codomains are finite powersets. If $\vfpl $ is such a
function, then $\vfpl\cons \pP\arr\Vset$ is the extension of $ \vfpl $  in
which $\pP$ is mapped onto $\Vset$,  assuming that $\pP$ is  not in $\dom
\vfpl $; conversely  $\resmin{\vfpl}{\pP}$ is the restriction of $\vfpl $ in
which the entry for $\pP$ is erased,
both in the domain and codomain of $\vfpl$. 
We also write
$\pP\arr\Vset,\qQ$ as  an
abbreviation for $\pP\arr (\Vset\cup \{\qQ\})$; and 
 $\vfpl \inDom \pP$
(resp.\ $\vfpl \NOTinDom \pP$) 
 means that $\vfpl $ is (resp.\ is not) defined on $\pP$. 
Function 
   $\vfpl[_1]\cap \vfpl[_2]$ has domain
 $\dom{\vfpl[_1]} \cap
   \dom{\vfpl[_2]}$, 
and then is defined 
as 
   $({\vfpl[_1]} \cap {\vfpl[_2]})(p) = \vfpl[_1](p) \cap
   \vfpl[_2](p)$; 
   and likewise for $\vfpl[_1]\cup\vfpl[_2]$.
   We write 
$\vfpl_1 \subseteq \vfpl_2$ if 
 $\dom{\vfpl[_1]} \subseteq 
   \dom{\vfpl[_2]}$, and for each $\pP \in \dom{\vfpl[_1]}  $, we have 
   ${\vfpl[_1]}(\pP) \subseteq {\vfpl[_1]}(\pP)$.

   \section{A Type System for Visibility}\label{s:types}

Our type system for visibility is built on top of that for sequentiality \cite{Hirschkoff_2021}.
While visibility could  be defined  as a standalone notion, the combination with
sequentiality makes the behavioural effects sharper and easier to grasp. 

Intuitively, sequentiality ensures us that at any time
  at most one interaction can occur; i.e., there is a single computation
  thread.  
A process that holds the thread, called \emph{active},
  decides what the next interaction can
  be.  It does so by offering a single particle (input or output) 
that controls the thread. 
A given name may exercise the control on the thread either in   output or in
input. Following~\cite{Hirschkoff_2021}, we require that, in higher-order names, the thread is carried
by outputs,
whereas, in first-order names, it is carried by inputs.
An input that carries the thread maintains the thread, whereas an output releases it. 
(These appear to be, practically, the
most interesting combinations; other combinations could be adopted, with the expected
modifications.) 
Thus, for instance, the following process $P$  is well-typed: 
\[ 
P \defi  \big( h(x).  \fbout a x {b} .  (    b(z,c').Q | \fout h x )  \big) ~|~  
\fout k v
\] 
The initial particles in $P$ are  the input at $h$ and the output at $k$; 
  however only the input at  $h$ carries the thread, as $h$ and $k$ are  first-order.
Underneath the input, the thread is maintained and released in the output at $a$ (the
thread will be acquired by the process in the external environment performing the input at
$a$); later, when the input at $b$ is performed,  process $Q$  obtains
the thread back. Sequentiality however does not 
 exclude non-determinism: e.g., an output particle
 may have the possibility of interacting
with different inputs. 


The type system for  \emph{visibility} intuitively allows   us to specify  which possible
higher-order outputs can be performed by a process that becomes active. 
Type environments are 
 partial functions, called 
 \emph{visibility functions}, acting on higher-order names and, for
 active processes, on the special token $\cur$, representing the
 active thread. (Later, when studying
 well-bracketing, type environments will have both a visibility
 function and a name stack.)

\begin{df}[Visibility function]
  A \emph{visibility function}
 is 
  a finite partial function
 $\vfpl: \Nho \cup \{\cur\}\to \mathcal{P}_{\tt fin}(\Nho)$.
Moreover the function should be 
  \emph{transitive}, that is, 
  for all $o\in \mathtt{dom}(\vfpl),$ if 
$\vfpl \inDom a$ and  
  $ a\in \vfpl(o)$, 
 then  $  \vfpl(a)\subseteq \vfpl(o)$.
\end{df}

To see the use of 
$\cur$, suppose
 $\vfpl (\cur) = \{a,b\}$;  this means that  
a $\vfpl$-process  is active, and may only perform outputs at $a$ and $b$,
as in the following process: 
\[ 
\fbout a v c. c(d). \boutDS b w .\nil
\]
Name $c$ is created in the output at $a$ and therefore $c$ inherits
$a$'s visibility.

Similarly, if $\vfpl (d) =  \{a,b\}$  then, in a 
$\vfpl$-process
$d(x,c). P $, the continuation  $P$ of the  input at $d$ 
may only perform outputs at $a$, $b$ or at  $c$ (as $c$ is the
higher-order name received in the input).
The use of visibility functions also allows us to sort out implicit
form of higher order store; in the following example, $p,q$ are higher
order names used for continuations. 

\begin{ex}[Implicit forms of higher-order store]
  In the $\lambda$-calculus, if a term without higher-order references
  is well-bracketed then it necessarily obeys visibility. In our calculus,
  even though we do not have explicit higher-order references, we
  can simulate higher-order store. Consider
  \[ P\defi\out{a}{}{b,p}.b(n,c,q).\out{q}{n}{}.p(m).\out{c}{}{}.\nil
    \]
  In $P$ the visibility $\vfpl(p)$ is defined when the output at $a$
  is made, therefore $c\notin \vfpl(p)$, and the final output at $c$ breaks
  visibility; indeed, the input at 
  $b$, which justifies the output at $c$, is answered by the
  output at $q$ before the output at $c$. 
  %

\end{ex}



Transitivity ensures us that the visibility functions, 
 viewed as `named sets', 
represent transitive sets.
Transitivity is needed to guarantee the behavioural expectations of
visibility.  If a process $P$ 
 calls a service $a$, with actual
(higher-order) parameter $b$, then an output triggered by such a call
(and possibly directed back to $P$)
should be either at $b$ or at a name in the visibility of $a$;
however,  the server $a$ might call an external service $c$; this call
should not allow an increase of the visibility associated
to  $a$; 
hence the transitivity constraint that the
visibility of $c$  is contained in that of $a$. 
Technically, transitivity is necessary  to ensure preservation of
typing under reduction (see Theorem~\ref{t:SR1} and Example~\ref{ex:tr}).

\begin{rem}[Transitivity and functional languages]
\label{r:trans:lambda} 
The issue of transitivity does not arise 
in the semantics of purely functional languages such as the
$\lambda$-calculus \cite{10.1007/978-3-030-72019-3_13}: visibility
functions do appear, but the domain and codomain of such functions are
disjoint (technically, in the game semantics terminology, 
the domain contains names introduced by 
Player, whereas the codomain contains names introduced by Opponent). This 
 makes
transitivity trivial. 
\end{rem}

In the typing of a parallel composition, the visibility function has
to be split between the two components of the composition. 
This is important for elements of the domain of the function that are
linear, and therefore may only appear in one, but not both,
components; this is the case for 
$\cur$, and the  linear continuation names that will be introduced in
Section~\ref{s:wb}.

\begin{df}[Split]
  Let $\vfpl$ be a visibility function, we define 
  $\Bsplit{\vfpl}$ as the set of pairs $(\vfpl[_1], \vfpl[_2])$
  such that
$\vfpl[_1] \cup \vfpl[_2] = \vfpl$ and $\cur \notin \dom{\vfpl[_1]}\cap \dom{\vfpl[_2]}$. 
\end{df}%

\begin{figure}[t]
 \begin{mathpar}
 \prftree[l]{\rname{O\sdash HO}}{a\in \vfpl(\cur)}
      {\typ{\typpl{
\resmin{\vfpl}{\cur}
 \cons b\arr \vfpl(\cur),b }{\ci{p},\stack}}{P}}{\typ{\typpl{\vfpl}{\stack}}{ \out{a}{e}{b}.P}}
\and
 \prftree[l]{\rname{I\sdash HO}}
      {\typ{ \typpl{\vfpl\cons  \cur \arr \vfpl(a),b}{}}{ P} }
      {\typ{ \vfpl}{ \inp a{ x,b} . P }}
\and
\prftree[l]{\rname{O\sdash FO}}
{
\typ{\typpl{\vfpl}{\emptyset}}{P }}{
\vfpl \NOTinDom \cur
}
{\typ{\typpl{\vfpl}{\emptyset}}{ \fout{h}{e}.{P}}}
      \and
      \prftree[l]{\rname{I\sdash FO}}
{
\typ{\typpl{\vfpl}{\emptyset}}{P }}{
\vfpl \inDom \cur
}
{\typ{\typpl{\vfpl}{\emptyset}}{  h(x).P}}
\and
\prftree[l]{\rname{REP}}
{
\typ{\typpl{\vfpl}{\emptyset}}{\alpha . P }}{
\vfpl \NOTinDom \cur
}
{\typ{\typpl{\vfpl}{\emptyset}}{!  \alpha .P}}
\and
\prftree[l]{\rname{SUCC}}
{
}{
}
{\typ{\typpl{\vfpl}{\emptyset}}{  \omega.P}}
\and
  \prftree[l]{$\mathtt{RES\sdash HO}$}{\typ{\typpl{\vfpl}{\stack}}{P}}
  {\typ{\typpl{\resmin\vfpl a}{\stack}}{\nu a P}}   
\and
 \prftree[l]{\rname{RES\sdash FO}}{\typ{\typpl{\vfpl}{\stack}}{
     P}}{\typ{\typpl{\vfpl}{\stack}}{ \nu h P}}
\and
\prftree[l]{\rname{NIL}}
{\vfpl \NOTinDom \cur
}{
}
{\typ{\typpl{\vfpl}{\emptyset}}{  \nil}}
\and
\prftree[l]{\rname{SUM}}
{\typ{\typpl{\vfpl}{\emptyset}}{ P_1}
\andalso \typ{\typpl{\vfpl}{\emptyset}}{ P_2}
}{
}
{\typ{\typpl{\vfpl}{\emptyset}}{ P_1+ P_2}}
\and
            \prftree[l]{\rname{PAR}}
            {\typ{\typpl{\vfpl_1}{\stack[_1]}}{P_1}}{\quad\typ{\typpl{\vfpl_2}{\stack[_2]}}{
          P_2}}{\quad (\vfpl_1,\vfpl_2)\in\Bsplit{\vfpl}}
{\typ{\typpl{\vfpl}{\stack}}{ P_1|P_2}}
\and
\prftree[l]{\rname{IF}}
{\typ{\typpl{\vfpl}{\emptyset}}{ P_1}
\quad \typ{\typpl{\vfpl}{\emptyset}}{P_2}
}{
}
{\typ{\typpl{\vfpl}{\emptyset}}{ \ifte ef{P_1}{P_2}}}
 \end{mathpar}
 \caption{Typing rules for visibility}
  \label{fig:typ}
\end{figure}

The typing rules are given in Table~\ref{fig:typ}. We comment on a
few of them.
Rule \rname{O\sdash HO} specifies that an output at a \Nho name $a$ owns the
thread, and that name $a$ must be in the current visibility. 
In the
continuation $P$ (which is not active) the visibility for $b$
is determined by the current visibility $\vfpl(\cur)$ 
together with $b$ itself, to allow for
recursive calls at $b$.
Symmetrically, in rule \rname{I\sdash HO}, a process performing an input at $a$ acquires
the thread, with a visibility given by $\vfpl(a)$ plus the received
name $b$. 
A replicated input is well-typed only if it is not active (rule \rname{REP}).
When typing parallel composition, we use \texttt{split} 
  to insure that
at most one of the components holds the thread.
We simply remove the name from $\vfpl$ when typing a restriction (rule
\rname{RES\sdash HO}). 
Visibility plays no role in rules \rname{I\sdash FO} and
\rname{O\sdash FO};
the only checks made are given
by sequentiality, concerning the respect of the  thread. 

\begin{ex}
Consider 
\[P \defi 
\begin{array}[t]{l}
\; a(x,b). \fbout b {x+1} d . a(x',b').  \fbout b
{x+x'}{ d'}.\nil  \\
| 
\; a(x_1,b_1). \fbout {b_1}{x_1}{d_1} . d_1(d_2). 
\fbout c {x+1}{c_1}. \nil
\end{array}
\]
The process has three inputs at $a$; the first two are sequential, and
make no requirements  on visibility, as their continuations only
perform outputs at  names received in the input; the third one, in
addition, has the effect of creating 
a name $d_1$, whose input may activate an output at
the free name $c$. This input requires that in the  visibility $\Delta $ 
of $P$, we find $c \in \Delta (a)$.

\end{ex}
 
\begin{ex}
\label{exa:notTypOut}
(In this example, we ignore the first-order components of actions, as
they are not relevant.)
A process of the form
\[ P \defi \boutDS a b . (Q| c(d). \boutDS b {d'}. R )\]
is not typable:  the 
initial visibility of $c$ may not include $b$, which is created fresh
by  $a$; yet, $P$ may perform an output at $b$ 
after the input at
$c$. Indeed, while $P$ does not directly create a form of
higher-order store, it may do so indirectly: the external
environment, after completing the treatment of the initial call at $a$,
is supposed to `forget' the higher-order name $b$.  However, this is
not the case, as the environment may 
regain access to $b $ by invoking the (unrelated) name $c$.
In contrast $P$  would become typable 
(in the straightforward extension of our type system with polyadicity)
if the initial output at $a$ would be 
replaced by  
$\boutDS a {b,c}$: now $c$ 
becomes
a fresh name, initially unknown, and created in the initial output,
together with $b$, so that the visibility for $b$ and $c$ can be the
same and can include both names.  
\end{ex}  

The following lemma shows that a visibility function may always be
enlarged (both in the domain and codomain), as long as sequentiality
(i.e., being defined on $\cur$) is respected.  

\begin{lm}[Weakening]
   \label{lm:wk}
 Suppose     $\typ{\vfpl[_1]}{P}$,
     $\vfpl[_1]\subseteq\vfpl[_2]$, and ($\vfpl[_2]\inDom\cur
$ iff $ \vfpl[_1]\inDom\cur$).
Then  $\typ{\vfpl[_2]}{P}$.
\end{lm}
Some other technical lemmas that are used in proofs are presented in Appendix~\ref{sec:typlm}.



\begin{thm}[Subject reduction]
\label{t:SR1}
Suppose $\typ{\typpl{\vfpl}{\stack}}{ P}$ and 
  $P\trans{\mu}P'$. We have:
  \begin{itemize}
  \item if $\mu$ is a $\tau$ action or a first-order input, then
    $\typ{\typpl{{\vfpl}{}}{\stack['']}}{P'}$; the same holds if $\mu$
    is a first-order output and
    $\vfpl\NOTinDom\cur$; 
  \item if $\mu = \out{a}{v}{b}$ then $a\in \vfpl(\cur)$ and 
$\typ{\typpl{ \resmin{\vfpl}{\cur}\cons b\arr \vfpl(\cur),b}{\ci{p},\stack}}{P'}$;
  \item if $\mu=\inpA{a}{v}{b}$ and $\vfpl\NOTinDom\cur$, then $\typ{\typpl{ \vfpl \cons \cur\arr \vfpl(a),b}{\co{p},\stack}}{P'}$.
  \end{itemize}
\end{thm}


\begin{ex}
  \label{ex:tr}
  Transitivity of visibility functions is necessary.
  Indeed, consider process $P \defi a.\out{b}{}{} .Q | \out{a}{}{}$, and
the \emph{non-transitive} function  $\vfpl =
  a\arr \{a,b\}, \cur \arr a$. 
We  have $\typ{\typpl\vfpl{}}P$.
However $P \trans{\tau} \out{b}{}{}.Q$, with  $\typpl\vfpl{}\nvdash
\out b{}{}.Q$, violating
  Subject
  Reduction (Theorem~\ref{t:SR1}).
\end{ex}


To adapt
the generic definitions of behavioural equivalence in 
 Section~\ref{s:lang} to 
the type system for visibility,
 we only have to specify the meaning of compatibility
between typing environments (which in our case are visibility functions).
Compatibility means ensuring that processes typed under the two typing environments
cannot both be active.

\begin{df}
  \label{d:compatible}
   Two typings $\vfpl[_1]$ and $\vfpl[_2]$ are \emph{compatible} if $\cur \notin \dom{\vfpl[_1]}\cap \dom{\vfpl[_2]}$.
\end{df}

We discuss some examples of  equivalences  in which  the behavioural constraints
imposed on legal observers by visibility are essential:  the
equivalences would  otherwise be broken.

\begin{ex}
\label{ex:visBehEffect}
Consider a process
\[
P\defi \boutDS a{b,c}. 
P' \andalso   \mbox{ for }\andalso P' \defi 
( b(\uuuu).R  ) 
|
 c(c_1). \boutDS {a} {b',c'} . c'(c_2) . \boutDS b \uuuup
. Q  
 \]
where we assume $b$ does not occur in $ Q$ and $R$.
Process $P$ interrogates an external service $a$, providing access to
 local services $b$ and $c$, and the derivative is $P'$.  
Now, $P'$ may  either be interrogated at $b$, 
or at $c$. In the former case, the output 
$ \boutDS b \uuuup
. Q$ becomes dead code (as the environment may not install any input at
$b$). In the latter case, 
the environment is again called at $a$; now,  
since  visibility ensures us the environment 
cannot remember $b$,  the input and the output 
 at $b$ in $P'$ may only   interact with each other;  
we can therefore inline the call at $b$. 
We can express this behaviour
using  expansion as follows: 
\[P \,\conteqNOtype\,    \boutDS a{b,c}. \Big( 
\begin{array}[t]{ll}
&b(\uuuu).(R | 
 c(c_1). \boutDS {a} {b',c'} . c'(c_2) . \nilact )
 \\
+&  c(c_1). \boutDS {a} {b',c'} . c'(c_2) .
 \tau . \res \uuuup ( Q| R\sub \uuuup \uuuu) 
\Big) \end{array}
  \] 
where $\nilact$ indicates an active process that cannot perform any
action (e.g.,  $\res b_2 ( \out{b_2}{}{}.\nil)$).
(This equality holds under any visibility, hence we use the symbol $\conteqNOtype$.) 

If we replaced the second output at $a $ with  an output at a
different free name, say $a'$, then the processes could still be
typable, and the equalities above would still hold. 
However, the equalities would not hold if we replaced 
the second output at $a$ with an output
at $c_1$. Indeed, 
while the environment may not store $b$ or $c$ for using them in a
later call at $a$ (or at an unrelated  name such as $a'$), 
the environment is allowed to use $b$ and $c$ within services that are
created following the first call at $a$; for instance, the
environment  could be 
\[ 
a(b,c). \boutDS c{c_1} . ! c_1 (c_1'). \boutDS b{d''}. \nil
\]



\end{ex}

The following example illustrates the use of 
 visibility  to perform  garbage-collection on 
  names that will  never be used. 
We refer  to Appendix~\ref{a:deadcode} for the full  definition of  the corresponding 
 pruning
function.

\begin{ex}
\label{ex:prun}
Consider a typing environment $\vfpl$, a $ \vfpl$-process
  \[P \defi !a .Q_1 \ \big|\  !b. \out{c}{}{}.!a.Q_2\  \big|\  !d.\out{a}{}{}.\nil\]
  and a visibility $\vfopp$ compatible with $\vfpl$ such that $\vfopp
  \cup \vfpl$ is transitive.
  $\vfopp$ is intended here to represent the observer's visibility.
  Suppose $a\notin \vfopp(\cur)$; that
  is, the observer is active but its current visibility does not include $a$ (which may
  however be in the visibility of other names in $ \vfopp $).  We then have
  $P \conteq \vfopp \vfpl !b.\out{c}{}{}.\nil$.  The pruning is possible because, intuitively,
  the transitivity of $\vfpl \cup \vfopp$ ensures us that an observer, whenever
  active, cannot  have $a$ in its current visibility.  For instance, if the observer interrogates $b$, then
the   transitivity of the visibility functions  ensures us  that $a\notin \vfpl(b)$;
 then $c\in \vfpl(b)$ (which is necessary for $P$ to be typable)  gives us  also
  $a\notin \vfopp(c)$. 
Similarly, exploiting transitivity,  we deduce that 
also 
$d\notin \vfopp(\cur)$, and that an observer may never be able  to interrogate $d$; 
thus we can also remove
  the input on $d$.
\end{ex}

\section{Labelled Semantics}
\label{s:labels}

\subsection{Trace-based proof techniques}
\label{ss:traces}

\iflong
\else
The definition of may-testing makes use of
contextual quantifications. 
We study here proof techniques
without such quantifications, as forms of  trace equivalence.
For lack of space, analogous proof techniques for barbed equivalence are only  reported in
Appendix~\ref{sec:lab_bisim}.
\fi
The first step is to express 
the   transitions for a process that are allowed by an observer subject
to certain visibility constraints. We call these 
the  \emph{type-allowed transitions}.
They
are  
of the form 
$\typtrans{\typopp{\vfopp}{}}{P}{\mu}{\typopp{\vfopp[']}{}}{P'}$, to
be read `an observer that has visibility $\vfopp$ 
allows  process $P$ to make a transition $\mu$ and, as a result of the
transition, the process  becomes $P'$ and the visibility of the
observer becomes  
$\vfopp[']$'.
Type-allowed transitions
are defined by the rules in Figure~\ref{fig:vlts}.
Intuitively, an observer typable under $\vfopp $ can initially test
inputs of processes at names in $\vfopp (\cur) $, if $\vfopp 
 \inDom \cur$, 
and 
 outputs of
processes in $\dom\vfopp $ otherwise.
The weak transitions $\Longrightarrow $ and  $\Arr\mu$, 
are defined in the usual manner from the strong one; 
for instance, 
 $\typTRANS{\typopp{\vfopp}{}}{P}{\mu}{ \typopp{\vfopp[']}{}}{P'}$
holds if 
 $\typRed{\typopp{\vfopp}{}}{P}{ \typopp{\vfopp}{}}{P''}$,  and 
$\typtrans{\typopp{\vfopp}{}}{P''}{\mu}{\typopp{\vfopp[']}{}}{P'''}$,
and 
 $\typRed{\typopp{\vfopp[']}{}}{P'''}{ \typopp{\vfopp[']}{}}{P'}$, for some 
 $P''$ and $P'''$.

\begin{figure}[t]
{\small 
  \begin{mathpar}
    \prftree[l]{\rname{TrO\sdash HO}}{P\xrightarrow{\out{a}{v}{b}}P'}{a\in
      \mathtt{dom}(\vfopp)}{\vfopp\NOTinDom\cur}{\typtrans{\typopp{\vfopp}{}}{P}{\out{a}{v}{b}}{\typopp{\vfopp\cons
          \cur \arr \vfopp(a),b}{}}{ P'}}
\and
  \prftree[l]{\rname{TrI\sdash HO}}{P\trans{\inpA avb}P'}{a\in
      \vfopp(\cur)}{\typtrans{\typopp{\vfopp
        }{}}{P}{\inpA avb}{\typopp{\resmin\vfopp\cur \cons b\arr
          \vfopp(\cur),b}{}}{ P'}}
%
%
\and
  \prftree[l]{\rname{TrO\sdash FO}}{P \trans{\out{h}{v}{}} P'}{\vfopp\inDom\cur}{\typtrans{\typopp{\vfopp}{}}{ P}{\out{h}{v}{}
      }{ \typopp{\vfopp}{}}{ P'}}
\and
  \prftree[l]{\rname{TrI{\sdash FO}}}{P \xrightarrow{\inpACT hv} P'}{\vfopp\NOTinDom\cur}{\typtrans{\typopp{\vfopp}{}}{ P}{\inpACT hv }{
      \typopp{\vfopp}{}}{ P'}}
  \and
   \prftree[l]{\rname{TrTau}}{P\trans{\tau} P'}{\vfopp\NOTinDom\cur}{\typtrans{\typopp{\vfopp}{}}{P}{\tau}{\typopp{\vfopp}{}}{
        P'}} 
  \end{mathpar}
}  \vskip -.3cm 
 \caption{Visible LTS}
  \label{fig:vlts}
\end{figure}

An action $\tau$ is called a \emph{reduction} (or an \emph{internal action}); the
remaining actions are   
 called
\emph{interactions}.
We use \trace{} to range over \emph{traces}, which are finite
sequences of
interactions 
$\mu_1\ldots\mu_n$ ($n \geq 0$)
in which the bound names are all distinct and different from the free names 
(the definition of bound and free names is as for processes, viewing a trace as a sequence
of prefixes).
A trace $\mu_1\ldots\mu_n$
is 
\emph{trace of $P_0$ under visibility $\vfopp[_0]$} 
if there are $P_i,\vfopp[_i]$ with 
 $\typTRANS{\vfopp[_i]}{P_i}{\mu_i}{\vfopp[_{i+1}]}{P_{i+1}}$, for 
 $0\leq i< n$.

\begin{df}[Typed trace equivalence]
Two $\vfpl$-processes $P$ and $Q$ are 
\emph{trace equivalent at $\vfopp$}, written 
$P \trEqV \vfpl \vfopp Q$, if $\vfpl$, $\vfopp$ are compatible
 and $P$, $Q$ have the same sets of  
traces  under  $\vfopp$.
\end{df}

We show the main compositionality lemmas, and a trace-based
characterisation of 
 may testing. Other technical lemmas needed in proofs may
be found in Appendix \ref{sec:treqlm}.

\begin{lm}[Restriction]
  \label{lm:res}
 If 
$P \trEqV \vfpl \vfopp Q$
then also 
$\res a P \trEqV{\resmin{\vfpl}{a}}{ \resmin{\vfopp}{{a}}} \res a Q$.
\end{lm}
\begin{lm}[Input]
  If $P \trEqV \vfpl \vfopp Q$ and
$\vfopp \NOTinDom \cur$; 
then also $a(x,b) .P \trEqV{\resmin{\vfpl}{\cur,b}}{ (\resmin{\vfopp\cons \cur \arr \vfopp(b))}{b}} a(x,b).Q$.
\end{lm}
\begin{lm}[Output]
  If $P \trEqV \vfpl \vfopp Q$ and
$\vfopp \NOTinDom \cur$, 
then also $\out{a}{e}{b}. P \trEqV{ (\resmin{\vfpl\cons \cur \arr \vfpl(b))}{b}}{ \resmin{\vfopp}{\cur,b}} \out{a}{e}{b}.Q$.
\end{lm}

\begin{lm}[Parallel composition]
  \label{lm:parcomp_tr}
Suppose $P \trEqV \vfpl \vfopp Q$
and 
  $(\vfopp[_1], \vfopp[_2])\in \mathtt{split}(\vfopp)$. 
Then 
  $\typ{\typpl{\vfopp[_2]}{}}{R}$
implies
 $P| R \trEqV{(\vfpl \cup \vfopp[_2])^+} {\vfopp_1} Q|R$ where $(-)^+$ indicates the transitive closure.
\end{lm}
In  the proof of Lemma~\ref{lm:parcomp_tr},
a trace of $P|R$ is split into
a trace for $P$ and one for $R$. 
We have then to ensure that the condition of the lemma are
maintained throughout the trace, notwithstanding  modifications of
the visibility functions.  Some care is needed in case of
synchronisations between $P$ and $R$, where the typing of the  external
observer has to remain the same; we also use 
the restriction Lemma~\ref{lm:res}.
Other cases also exploit the extension Lemma~\ref{l:tr_ext}, and
the transitivity   of the visibility function, to make sure that 
the views on free names are not increased along the trace.

\begin{thm}[Characterisation of may testing using trace equivalence]
   Suppose $\typ{\vfpl}{P,Q}$ then $P \trEqV \vfpl \vfopp Q$ if and only if $P\conteq{\vfpl}{\vfopp} Q$.
\end{thm}

The sketch of the proofs is the same as for characterisations of may testing with trace
equivalence in untyped calculi, exploiting substitutivity properties of trace equivalence
in the case of Soundness, and 
reason by contradiction in the case of 
Completeness. In addition, one has to take care of the observer visibilities, how it
changes, and use type-allowed transitions.
A similar results of characterisation also hold for
barbed equivalence, using a typed form labelled  bisimilarity, and  are presented in Appendix~\ref{sec:lab_bisim}.


We can use typed trace equivalence to prove equalities such as those in
Examples~\ref{ex:visBehEffect} and \ref{ex:prun};
 the proofs are simple, following the definition of
the equivalence and exploiting
 type-allowed transitions. 
See also Appendix~\ref{a:deadcode} for the proof about the pruning function
that generalises Example~\ref{ex:prun}.
 (Similarly, the proofs can be carried out for
barbed equivalence, using the labelled bisimilarity in
Appendix~\ref{sec:lab_bisim}.)   


\subsection{Visibility property on traces}\label{s:view}
We formalise the link between our type system and the notion of
visibility in game semantics~\cite{10.1007/978-3-030-72019-3_13}.
The
reader not familiar with game semantics may safely skip the section.
  In game semantics, the notions of visibility and well-bracketing
  are defined based on a justification relation between moves
  performed by Player. These moves correspond to interactions on
  higher-order names in our setting, and the notion of justification
  can be formulated as follows.
\begin{df}
  Let $\mu_1,\mu_2$ be two interactions at \Nho names. We say that
  \emph{$\mu_1$ justifies $\mu_2$}, noted
  $\mu_1 \justif \mu_2$,
  if:
  $(i)$ either $\mu_1 = \inpA{a}{v}{b}$ and $\mu_2 = \out{b}{v'}{c}$,
 or $(ii)$ $\mu_1 = \out{a}{v}{b}$ and $\mu_2 = \inpA{b}{v'}{c}$.
  When this is not the case, we write $\mu_1\njustif\mu_2$.
\end{df}
$\mu_1\justif\mu_2$ captures the dependency induced by the fact that
$\mu_2$ performs an interaction at a name that has been previously
introduced by $\mu_1$. In contrast with the situation in game
semantics, though, there is no notion of question and answer in our
setting.

We can now define the notion of view.
%
%
In the definition below, we allow a starting
visibility \Vset,  because we want to consider processes that
already have some knowledge about the observer (Opponent).

\begin{df}[View]\label{d:view}
  Given 
  $\Vset\subseteq
  \Nho$, we define the \emph{view with starting visibility $\Vset$ of
    a
    trace \trace}, written
  \view\Vset\trace, as follows:
    \begin{mathpar}
  \mathtt{view}_\Vset(\varepsilon) = \Vset
  \and
   \mathtt{view}_\Vset(\trace_1\mu_1\trace_2\mu_2 )=
    \mathtt{view}_\Vset(\trace_1)\cup
    \fNho{\mu_1} \cup \fNho{\mu_2}
    \text{ if } {\mu_1}\justif{\mu_2}
    \and
     \mathtt{view}_\Vset(\mu_0\dots\mu_n) = \Vset\cup \fNho{\mu_n}
    \text{ if } \forall 0\leq i\leq n-1, {\mu_i}\njustif{\mu_n} 
    \end{mathpar}
    where $\fNho{\mu}$ denotes the set of higher-order names appearing
  in the action $\mu$.
\end{df}

Views capture the idea 
that when an action is justified by another
one, the names created in between 
must have been forgotten,
otherwise this would mean that the process had the ability to store these
names in order to reuse them.
%
In the last clause in Definition~\ref{d:view},
the action is justified by an action that happened before the
beginning of the trace, therefore it has
the starting visibility.
In particular, \view\emptyset\trace{} yields
the usual definition in game semantics. 
\begin{df}[Respecting views]\label{d:respect:view}
  Consider $\Vset\subseteq \Nho$, a trace $(\mu_0\dots\mu_n)$ \emph{respect views under starting visibility }$\Vset$ if $\mathtt{fn}(\mu_0) \subseteq \Vset$ and $\forall 1\leq i\leq n, \mathtt{fn}(\mu_i)\subseteq \mathtt{view}_\Vset(\mu_0\dots\mu_{i-1})$
\end{df}
\begin{lm}\label{l:view}
  Consider $\vfpl,\vfopp,P$ such that $\typ{\vfpl}{P}$, $\vfopp$ is
  compatible with $\vfpl$, and $\vfpl \cup \vfopp$ is
  transitive. Let $\Vset=(\vfopp \cup \vfpl)(\cur)$, 
  then every
  trace of $P$ under visibility $\vfopp$ respects views with starting
  visibility \Vset.
\end{lm}

The idea of the proof is that for a trace $(\mu_0\dots\mu_n)$ there
exist $P_0,\dots,P_{n+1}$, $\vfpl[_0],\dots,\vfpl[_{n}]$,
$\vfopp[_0],\dots,\vfopp[_{n+1}]$ such that $P_0 = P$, $\vfpl[_0] =
\vfpl$, $\vfopp[_0] = \vfopp$ and for all $0\leq i \leq n$,
$\typtrans{\vfopp[_i]}{P_i}{\mu_i}{\vfopp[_{i+1}]}{P_{i+1}}$ and
$\vfpl[_i] \vdash P_i$.
We show by induction that for each $1\leq i \leq n$, $(\vfopp_i \cup
\vfpl_i)(\cur) \subseteq \mathtt{view}_\Vset(\mu_0\dots \mu_{i-1})$; it thus
 follows that the trace respects views under starting visibility
\Vset. The condition of $\vfpl\cup \vfopp$ being transitive is crucial, otherwise $(\vfopp_i \cup \vfpl_i) (\cur)$ could grow in unwanted ways.

\renewcommand{\typpl}[2]{\langle
 \textcolor{green}{#1}
 |\textcolor{green} {#2} \rangle
}
\renewcommand{\typopp}[2]{[
  \textcolor{red}{#1} |
  \textcolor{red}{#2}] 
}
\renewcommand{\typtrans}[5]{{#1}{#2} \xrightarrow{#3}
  {#4}{#5}}
\renewcommand{\typTRANS}[5]{{#1}{#2} \xRightarrow{{#3}}{#4}{#5}}
\renewcommand{\typTrans}[5]{{#1}{#2} \xRightarrow{\hat{#3}}{#4}{#5}}

\newcommand{\vispi}{\ensuremath{V\pi}}
\newcommand{\wbvispi}{\ensuremath{V\pi^{\mathrm{wb}}}}

\section{Adding Well-Bracketing}\label{s:wb}


\subsection{Type system}

In the previous sections we have studied 
 visibility in connection with  sequentiality. 
In this section we do the same for a refinement of 
 sequentiality,  namely well-bracketing.  
We begin by reviewing 
 well-bracketing, following~\cite{Hirschkoff_2021}, and 
 then outline how visibility is accommodated.

{\bf From sequentiality to well-bracketing.}
In
 languages without  control operators, well-bracketing intuitively means
that return-call interactions among terms follow a stack-based discipline.
In a well-bracketed system, 
 a service that is interrogated, say $S$, 
 acquires the thread  and is supposed to return a final result 
(unless the computation diverges)  thus releasing the thread. 
During its computation, $S$ may however  interrogate another service, say $T$,
which, upon completion of its computation, will return the result to
$S$; service $S$ must receive the answer from $T$ before returning the
final result. 
The implementation of this policy requires 
 \emph{continuation names}, that are used to handle the return
 messages.   When a service is interrogated, it receives a
fresh continuation  name that is used once, in output,  to deliver the
final result. Accordingly, 
a client that has interrogated the service will use
the continuation name in input to receive the result. 
 Thus continuation
names are \emph{linear}~\cite{DBLP:journals/toplas/KobayashiPT99}~---
they may only be used once. 
In contrast with~\cite{Hirschkoff_2021}, we do not require
\emph{receptiveness}  \cite{DBLP:journals/tcs/Sangiorgi99}~---
the  input-end of  the name needs not  be made available  as soon as
the name is created. Enforcing only linearity allows us to gain
flexibility. 
%
%
%
%
To ensure the well-bracketing policy, the type
system in~\cite{prebet:tel-03920089}  makes
use of stacks.
\begin{df}[\cite{prebet:tel-03920089}]
  A stack, $\stack$,  is a sequence of input- and output- tagged continuation names, in which the input and output tags alternate:
  \begin{mathpar}
    \stack ::= \stack[_\mathtt{O}] \sep \stack[_\mathtt{I}]

    \stack[_\mathtt{O}] ::= \co{p}, \stack[_\mathtt{I}] \sep \emptyset

    \stack[_\mathtt{I}] ::= \ci{p}, \stack[_\mathtt{O}] \sep \emptyset
  \end{mathpar}
  Moreover: a name may appear at most once with a given tag; and, if a name appears with both tags, then the input
  occurrence should immediately follow the output occurrence.
\end{df}

We write  $\co{p}\in \stack$ (resp.\ $\ci{p}\in \stack$)
if there is an occurrence of $\co{p}$ (resp.\ $\ci p$) in $\stack$.
We write $p\in \stack$ if either $\co{p}\in \stack$ or $\ci{p}\in
\stack$.

Intuitively,  a stack expresses the expected usage of the free continuation names in a
process.  For instance, if 
$P$ can be typed using stack ${\Tag {p_1} \OO, \Tag {p_2} \II, \Tag {p_3} \OO, \Tag {p_3} \II,  \Tag {p_4} \OO}$,
%
then $p_1,..,p_4$ are the free continuation names in $P$; 
among these,  $p_1$ will be used first,  in an output; 
then $p_2$ will be used, in an input  interaction with the
environment.  $P$ possesses both
the output and the input capability on $p_3$, so it shall perform a
reduction at $p_3$;
the computation for $P$ terminates with an output at $p_4$.
 This behaviour however concerns only the free continuation names of $P$: at any time
 when an output usage is expected, $P$ may decide to create a new continuation name and
 send it out, maintaining its input end. 


{\bf Visibility.}
We now discuss how to combine the visibility in
Section~\ref{s:types} with well-bracketing.
Higher-order  names are partitioned into
the set \nser{} of
\emph{server names}, and 
the set
 \ncon{}  of \emph{continuation
  names}. Now  $a,b,c$
range over server names, while  
 $p,q,r$   range over continuation
  names.
By convention, 
server names carry a triple consisting of a first-order value, a
server name and a continuation; whereas continuations carry a first-order
value and a server name. Accordingly,  outputs on higher-order names
are of the form
$\out{a}{n}{b,p}$ and $\out{p}{n}{b}$, and similarly for inputs.
First-order names  are as before.

Visibility functions are then
 transitive partial functions from
${\nser} \mathrel{\cup} {\ncon}\mathrel{\cup}{ \{\cur\}}$ to 
$\mathcal{P}_{\tt fin}({\nser}\mathrel{\cup}{\ncon})$.
Server names are handled essentially like \Nho
names in Section~\ref{s:types}. 
In contrast, some care is needed on continuation names because they
are linear. For this reason, some of the notations for handling
visibility functions have to be revisited. 
  A visibility function $\vfpl$ is \emph{duplicable}, written
  $\ser{\vfpl}$, if $\operatorname{dom}(\vfpl) \subseteq
  \nser$. This predicate is useful to insure that continuation
  names are used linearly.
  In particular, the definition of splitting is refined as follows:
  $\mathtt{split}(\vfpl)$ is the set of pairs $(\vfpl[_1], \vfpl[_2])$
  such that $ \vfpl[_1] \cup \vfpl[_2] = \vfpl$ and $\ser{(\vfpl[_1] \cap \vfpl[_2] )}$.

  Typing environments
include a stack;
%
thus a typing environment
is a pair $\typpl{\vfpl}{\stack}$ of
 a visibility function and a  stack.
  In the typing rule for parallel composition, we  rely on splitting
  as well as on
  $\mathtt{inter}(\stack[_1],\stack[_2])$, the interleaving of stacks
  $\stack[_1]$ and $\stack[_2]$. 
  
  \begin{df}[Interleaving]
We write $\inter{\sigma_1}{\sigma_2}{\sigma_3}$ if  
(i)  $\sigma_1$ is a
stack,  and (ii) $ \sigma $  is an interleaving of $\sigma_2$ and $\sigma_3$ as
by 
the following inductive rules:
\begin{enumerate}
	\item $\inter{\emptyset}{\emptyset}{\emptyset}$
	\item\label{inter:deux} $\inter{\Tag{p}{\OO},\sigma_1}{\sigma_2}
          {\Tag{p}{\OO},\sigma_3}$ if $\inter{\sigma_1}{\sigma_2}{\sigma_3}$
	\item\label{inter:trois} $\inter{\Tag{p}{\OO},\sigma_1}{\Tag{p}{\OO},\sigma_2}{\sigma_3}$ if $\inter{\sigma_1}{\sigma_2}{\sigma_3}$
	\item the same as~(\ref{inter:deux}) and~(\ref{inter:trois})
          with $\Tag{p}{\II}$ instead of $\Tag p\OO$
\end{enumerate}
\end{df}

\begin{figure}[t]
{\small
  \begin{mathpar}
    \prftree[l]{\rname{O\sdash ser}}{a\in \Vset}
      {\typ{\typpl{\vfpl\cons b\arr \Vset,b \cons p\arr \Vset,b}{\ci{p},\stack}}{P}}{\typ{\typpl{\vfpl\cons\cur \arr \Vset}{\stack}}{ \out{a}{e}{b,p}.P}}
      ~~
    \prftree[l]{\rname{O\sdash con}}{p\in \Vset}
    {\typ{\typpl{\vfpl\cons b\arr \Vset,b}{ \stack}}{
        P}}{\typ{\typpl{\vfpl \cons\cur \arr \Vset}{\co{p},\stack }}{
        \out{p}{e}{b}.P}} 
        \and
  \prftree[l]{\rname{I\sdash ser_1}}
      {\typ{\typpl{\vfpl\cons a \arr \Vset \cons \cur \arr \Vset,b,p }{\co{p},\stack}}{ P}}{\typ{\typpl{\vfpl \cons a\arr \Vset}{\stack}}{ a(x,b,p).P}}
      ~~
      \prftree[l]{\rname{I\sdash ser_2}}{\ser{\vfpl}}
      {\typ{\typpl{\vfpl\cons a \arr \Vset \cons \cur \arr \Vset,b,p }{\co{p}}}{ P}}{\typ{\typpl{\vfpl \cons a\arr \Vset}{\emptyset}}{ !a(x,b,p).P}}
    \and
    \prftree[l]{\rname{I\sdash con}}
  { \typ{\typpl{\vfpl\cons \cur \arr \Vset,b}{\co{q}}}{ P}}{\typ{\typpl{\vfpl \cons p\arr \Vset}{\ci{p},\co{q}}}{ p(x,b).P}}
  %
\and
        \prftree[l]{\rname{RES\sdash ser}}{\typ{\typpl{\vfpl}{\stack}}{P}}
        {\typ{\typpl{\resmin\vfpl a}{\stack}}{\nu a P}} 
      \and
        \prftree[l]{\rname{RES\sdash con_1}}{\typ{\typpl{\vfpl}{\stack[_1],
              \co{p},\ci{p},\stack[_2]}}{P}}
        {\typ{\typpl{\resmin\vfpl p}{\stack[_1],\stack[_2]}}{\nu p P}} 
        \and
      \prftree[l]{\rname{RES\sdash con_2}}{\typ{\typpl{\vfpl}{\stack}}{P}}
      {p\notin \stack}{\typ{\typpl{\resmin\vfpl p}{\stack}}{\nu p P}}
      %
      \and
      \prftree[l]{\rname{PAR}}
        {
          \begin{array}{ll}
      \typ{\typpl{\vfpl[_1]}{\stack[_1]}}{P_1} &
          (\vfpl[_1],\vfpl[_2])\in \Bsplit\vfpl
            \\ 
      \typ{\typpl{\vfpl[_2]}{\stack[_2]}}{P_2} &
            \stack\in \mathtt{inter}(\stack[_1],\stack[_2])
          \end{array}
        }
        {\typ{\typpl{\vfpl}{\stack}}{ P_1|P_2}} 
        \and
    \prftree[l]{\rname{NIL}}{\ser{\vfpl}}{\typ{\typpl{\vfpl}{\emptyset}}{ \nil}}
  \end{mathpar}
}    \vskip -.3cm\caption{Typing rules with visibility and stack
} 
  \label{fig:typ:wb}
\end{figure}

If a name
appears both in $\sigma_2$ and in $\sigma_3$ with the same tag, then
$\theinter{\sigma_2}{\sigma_3}$ is empty. 
Figure~\ref{fig:typ:wb} gives the typing rules. 
The
rules combine the handling of visibility
and the update of the stack.
For instance, in rule \rname{O\sdash SER}, newly created names $b$ and $p$
are added to the visibility when typing the continuation $P$, and $\ci
p$ is added on top of the stack, meaning that $P$ shall perform an
input at $p$ before interacting on other continuation names.
The output capability on $p$ is
transmitted along the output at $a$; correspondingly, in rules
\rname{I\sdash SER_1} and \rname{I\sdash SER_2}, $\co p$ is added to $\stack$.
In contrast with~\cite{Hirschkoff_2021}, we do not impose
$\stack=\emptyset$ in rule  $\mathtt{I\sdash SER_1}$, as we do not enforce input
receptiveness on continuation names.
%
Performing inputs and outputs at continuation names has the effect of
consuming the corresponding particle on top of the stack.
The stack is empty when typing a replicated input,
so to avoid duplication of  continuation names. 
%
%
%
In rules \rname{I\sdash SER_2} and \rname{NIL} we need to check that
$\vfpl$ is duplicable to make sure that all the input on continuations
are defined.   
    Rules \rname{SUCC},
  \rname{O\sdash FO}, \rname{I\sdash FO}, \rname{RES\sdash FO}, \rname{SUM} and \rname{IF} are like in
  Figure~\ref{fig:typ}, and are omitted.


\begin{thm}[Subject reduction]\label{t:SR:wb}
  Suppose $\typ{\typpl{\vfpl}{\stack}}{ P}$ and
  $P\xrightarrow{\mu}P'$. We have:
  \begin{itemize}
  \item if $\mu = \tau$, then either $\stack =
    \co{p},\ci{p},\stack['']$ with $p\in \mathtt{Ncon}\setminus
    \fnames{P'}$, in which case $\typ{\typpl{\resmin{\vfpl}{p}}{\stack['']}}{P'}$,

    or
    $\typ{\typpl{\vfpl}{\stack}}{P'}$ ;
  \item if $\mu = \out{a}{v}{b,p}$ then $a\in \vfpl(\cur)$ and
    $\typ{\typpl{\resmin\vfpl\cur\cons b\arr \vfpl(\cur),b\cons p\arr
        \vfpl(\cur),b}{\ci{p},\stack}}{P'}$ ;
  \item if $\mu=\out{p}{v}{b}$ then $\stack = \co{p},\stack['']$,
    $p\in \vfpl(\cur)$ and $\typ{\typpl{\resmin\vfpl\cur\cons b\arr
        \vfpl(\cur),b}{ \stack['']}}{P'}$ ;
  \item if $\mu=\inpA{a}{v}{b,p}$
    and $\vfpl\NOTinDom\cur$
    then $\typ{\typpl{ \vfpl \cons
        \cur\arr \vfpl(a),b,p}{\co{p},\stack}}{P'}$ ;
  \item if $\mu = \inpA{p}{v}{b}$
        and $\vfpl\NOTinDom\cur$
        then $\stack =
    \ci{p},\co{q},\stack['']$ and $\typ{\typpl{\resmin{\vfpl}{p} \cons
        \cur \arr \vfpl(p),b}{\co{q},\stack['']}}{P'}$ ;
  \item if $\mu = \inpACT hv$ or 
    ($\mu = \out{h}{v}{}$
         and $\vfpl\NOTinDom\cur$)
    then $\typ{\typpl{ \vfpl}{\stack}}{P'}$.
  \end{itemize}
\end{thm}


As previously, we have to define compatibility of typing environments so to be able
to use the contextual forms of behavioural equivalence from
Section~\ref{s:lang}.  

\begin{df}
  $\typpl{\vfpl[_1]}{\stack_1}$ and
  $\typpl{\vfpl[_2]}{\stack_2}$ are compatible
  if $\ser{(\vfpl[_1]\cap \vfpl[_2])}$
  and $\mathtt{inter}(\stack[_1],
  \stack[_2]) \neq \emptyset$. 
\end{df} 

\begin{ex}
\label{ex:INC}
Consider the process
\[
P\defi
\res h (
\begin{array}[t]{l}
 \outDS h 5 | \\
 \boutDS {a} {b,q} .  ( 
\begin{array}[t]{l}
 ! b(p). h(x). (\outDS h{x+1}.\nil| \outC p.\nil)  |  \\
 q. h(x). \outDS h{x} .   \boutDS c {d,r}. r .  h(y). \outDS h{y} . 
\ifte
  x y  {\outC s .\nil} \nilact 
  ))
\end{array} 
\end{array} 
 \]
Here $h$ is used as an integer reference, initially set to $5$.
The process calls a service $a$, giving access to a name $b$ that can be used to
increment the reference.  When the call at $a$  returns (at $q$),
the
reference is read twice and, in between, 
a call at $c$ is made. Without visibility, the two values read for $h$ might be
different: as $b$ has been exported, the environment could freely use $b$ and increment
the reference in between the two reads.  
Visibility ensures us that, when the initial call at $a$ is returned, name $b$ is
`forgotten'  by the environment, which cannot therefore use it anymore.    
Thus, if $Q$ is  the process defined like $P$ but with the last line replaced with 
 $q.   \boutDS c {d,r}. r . \outC s.\nil$, then we have $P  \conteqNOtype Q$. 
We refer to Appendix~\ref{a:reentrant:call} for an ML-like
presentation of the example above,  as well as for a longer example,
also showing how re-entrant calls may sometimes 
make  visibility boundaries hard to predict. 
\end{ex}

\subsection{Labelled semantics}

  To represent the point of view of the observer, we use typing
  environments of the form
  $\typopp{\vfopp}{\kcats}$, where $\vfopp$ is a
  visibility function and $\kcats$ is a \emph{stack}.
  Figure~\ref{fig:vlts:wb} defines the typed transitions, written
  $\typtrans{\typopp\vfopp\kcats}P\mu{\typopp{\vfopp[']}{\kcats[']}}{P'}$.
  Rules \rname{TrTau},
      \rname{TrO\sdash fo} and \rname{TrI\sdash fo} are similar to
      their counterpart in Figure~\ref{fig:vlts}, and are omitted.
\begin{figure}[t]
{\small
  \begin{mathpar}
     \prftree[l]{\rname{TrO\sdash ser}}{P\xrightarrow{\out{a}{v}{b,p}}P'}
    {a\in\dom\vfopp}
    {\vfopp\NOTinDom\cur}
    {\typtrans{\typopp{\vfopp}{ \kcats
        }}{P}{\out{a}{v}{b,p}}{\typopp{\vfopp\cons \cur \arr
          \vfopp(a),b,p}{ \co{p},\kcats }}{ P'}}
    \and
     \prftree[l]{\rname{TrO\sdash con}}{P\xrightarrow{\out{p}{v}{b}} P'
    }
    {p\in\dom\vfopp}
    {\vfopp\NOTinDom\cur}{\typtrans{\typopp{\vfopp}{\ci{p},\co{q},\kcats}}{P}{\out{p}{v}{
          b}}{\typopp{\resmin{\vfopp}{p}\cons\cur \arr \vfopp(p),b}{
          \co{q},\kcats }}{P'}}
    \and
  %
     \prftree[l]{\rname{TrI\sdash ser}}
    {P\trans{\inpA a{v}{b,p}}P'}
    {a\in \vfopp(\cur)}
    {\typtrans{\typopp{\vfopp
        }{\kcats}}{P}{\inpA{a}{v}{b,p}}{\typopp{\resmin\vfopp\cur
          \cons p\arr \vfopp(\cur),b \cons b\arr \vfopp(\cur),b}{
          \ci{p},\kcats}}{ P'}}
    \and
     \prftree[l]{\rname{TrI\sdash con}}
    {P\trans{\inpA p{v}{b}}P'}
    {p\in \vfopp(\cur)}
    {\typtrans{\typopp{\vfopp}{\co{p},\kcats}}{P}{\inpA p{v}{b}}{
        \typopp{\resmin\vfopp\cur\cons b\arr \vfopp(\cur),b}{
          \kcats }}{ P'}}
    \end{mathpar}
}  \vskip -.3cm
    \caption{Visible LTS with stack}
  \label{fig:vlts:wb}
\end{figure}

A trace $\mu_1\dots\mu_n$ is a \emph{well-bracketed trace of $P_0$ under $\typopp{\vfopp[_0]}{\kcats[_0]}$ } 
if there are $P_i,\vfopp[_i],\kcats[_i]$ ($1\leq i\leq n$) with 
 $\typTRANS{\typopp{\vfopp[_i]}{\kcats[_i]}}{P_i}{\mu_i}{\typopp{\vfopp[_{i+1}]}{\kcats[_{i+1}]}}{P_{i+1}}$
 for all $0\leq i< n$.

\begin{df}[Typed trace equivalence]
Two $\typpl{\vfpl}{\stack}$-processes $P$ and $Q$ are 
\emph{trace equivalent at $\typopp{\vfopp}{\kcats}$}, written 
$P \trEqV{\typpl\vfpl\stack}{\typopp{\vfopp}{\kcats}} Q$, if $\typpl{\vfpl}{\stack}$ and
$\typopp{\vfopp}{\kcats}$ are compatible and $P$ and $Q$ have the same
sets of   
well-bracketed traces  under  $\typopp{\vfopp}{\kcats}$.
\end{df}

This  typed version of trace equivalence enjoys properties, including substitutivity
properties, similar to those of the trace equivalence in Section~\ref{s:labels}.   
We only report the characterisation result with respect to may testing. 

\begin{thm}[Characterisation of may testing using trace equivalence]
   Suppose $\typ{\typpl{\vfpl}{\stack}}{P,Q}$; then
   $P \trEqV {\typpl{\vfpl}{\stack}}{\typopp{\vfopp}{\kcats}} Q$
   if and only if
 $P\conteq{\typpl{\vfpl}{\stack}}{\typopp{\vfopp}{\kcats}} Q$.
\end{thm}

The analogous result for typed bisimilarity is in
Appendix~\ref{sec:bis_wb}, together with 
 results about the shape of type-allowed traces akin to 
 those in  Section~\ref{s:view}.

\section{Conclusions  and Future Work}

In this paper we have studied the meaning 
 of visibility, originating from 
game-semantics
interpretations of
typed $\lambda$-calculi and reflecting absence of higher-order store, 
 in a calculus of pure
name-passing such as the $\pi$-calculus. 
In contrast with $\lambda$-calculi, in the $\pi$-calculus there is no
explicit notion of store, and no unique identities (e.g., the input
end of a name may have multiple occurrences,  arbitrarily
structured). These aspects determine considerable differences in the
semantic analysis of visibility. For instance, 
in the $\pi$-calculus
visibility has to be
enforced by means  of a dedicated type system; and the visibility
functions have to satisfy special properties such as transitivity. 
We have highlighted such differences, and investigated the
behavioural effects of visibility. 
In particular, we have shown full abstract characterisations of  may testing and
barbed equivalence, as form of  trace equivalence and labelled
bisimilarity.  

There are several directions for future work that we would like to
pursue. In game semantics,  further aspects related to first-order
 store have
been studied; for instance allowing   
 first-order references as values (that can be passed around)~\cite{DBLP:conf/icalp/Laird07,DBLP:conf/lics/JaberT16}, or
allowing to   store first-order references (i.e., references with 
 types such
 as {\tt int ref ref}~\cite{murawski_algorithmic_2018,kammar_monad_2017}). 
We would like to investigate the effects of such distinctions on our types
and proof techniques.

We have worked with \piI, a dialect of  $\pi$-calculus in which all
names exchanged are newly created. 
Thus \piI lacks the free-output construct, where the communication of 
 existing names is permitted.
As shown in the literature~\cite{DBLP:journals/entcs/Honda02,DBLP:journals/corr/abs-2011-05248,jaber:hal-03407123}, \piI is closer to game semantics than the
standard $\pi$-calculus. We would like to see whether our results can
be adapted so to accommodate also the free output construct. 
While free output can be modelled in \piI \cite{DBLP:journals/tcs/Boreale98},
it allows a direct representation 
of idioms  such as `delegate to
someone else the task of answering', or the expression of 
tail-call
recursion. 

We have studied visibility on top of type systems for sequentiality
and well-bracketing. Allowing concurrency would probably 
 require relaxing the
definition of
 the visibility functions; we leave the
details to future
work.

\bibliographystyle{plainurl}
\bibliography{bibliography/biblio.bib}

\appendix

\section{Section~\ref{s:lang}: Definition of the Operational Semantics}
\label{a:calc}


      \emph{Structural congruence}, written $\equiv$, is the smallest
      relation that is an equivalence relation, contains
      $\alpha$-conversion, is preserved by all operators of the
      calculus, and satisfies the following axioms:
  \begin{mathpar}
    \prftree{}{P | Q \ \equiv\  Q | P}
    \and
    \prftree{}{P | (Q | R) \ \equiv\  (P | Q) | R}
    \and
    \prftree{}{P|0 \equiv P}
    \and
    \prftree{}{P + Q\ \equiv \ Q +P}
    \and
    \prftree{}{P + (Q + R) \ \equiv\  (P + Q) + R}
    \and
    \prftree{}{P+0 \equiv P}
    \and  
    \prftree{a\notin \mathtt{fn}(Q)}{ (\new a P) | Q \ \equiv \ \new a
      (P |Q)}
    \and
    \prftree{}{\new a\new b P \equiv \new b\new a P}
    \and  
    \prftree{}{!\alpha.P \equiv !\alpha.P | \alpha.P}
  \end{mathpar}
Similar versions of axioms for restrictions of the form $\new h$
instead of $\new a$ have been omitted.

To define the labelled transition system (LTS), we introduce
\emph{actions}, defined by the following grammar:
\begin{mathpar}
  \mu\quad::=\quad \tau \midd \inpA a v{b} \midd \out avb\midd
  \inpACT h v\midd \out hv{}\midd\omega
\end{mathpar}
Actions are \emph{early} for first-order values, and \emph{late} for
the transmission of \Nho names. We
write $\bnames\mu$ (resp.\ $\fnames\mu$) for the bound (resp.\ free)
names of action $\mu$, defined by saying that 
$b$ is a {bound} name in actions $\inpA
avb$ and $\out avb$, and the other names in actions are {free}.

We suppose that any closed first-order expression $e$ can be evaluated to
some value $v$. We write \eval ev{} for the corresponding evaluation
relation. We furthermore suppose that equality between values can be
tested.

The LTS is defined, on closed processes, by the rules in Figure~\ref{f:lts}.

We write $P\bomega$ if $P\trans\omega$, i.e., if $P$ contains an
unguarded occurrence of an $\omega$ prefix at toplevel.
We write $P\nbomega$ when this is not the case.

\begin{figure}[t]
  \begin{mathpar}
    \prftree{}{a(x,b).P \trans{a\langle v\rangle (b)} P[v/x]}
    \and
    \prftree[]{\eval e v}
    {\out{a}{e}{b}.P \trans{\out{a}{v}{b}} P}
    \and
    \prftree{}{h(x).P \trans{h\langle v\rangle} P[v/x]}
    \and
    \prftree[]{\eval e v}
    {\out{h}{e}{}.P \trans{\out{h}{v}{}} P}
    \and
    \prftree{}{\omega.P \trans{\omega} \nil}
    \and
    \prftree{\alpha.P \trans{\mu} Q}{!\alpha. P \trans{\mu} Q | !\alpha . P}
    \and
    \prftree[r]{$a\notin \mathtt{fn}(\mu)$}{P \trans{\mu} Q}{\new a\ P \trans{\mu} \new a\ Q}
    \and
    \prftree[r]{$h\notin \mathtt{fn}(\mu)$}{P \trans{\mu} Q}{\new h\ P \trans{\mu} \new h\ Q}
    \and
    \prftree{ P \trans{\mu} P'}{P + Q \trans{\mu} P'}
    \and
    \prftree
    {P \trans{a\langle v\rangle (b)} P'}{Q \trans{\out{a}{v}{b}} Q'}{ P|Q \trans{\tau} \new b (P'|Q')}
    \and
    \prftree{P \trans{h\langle v\rangle} P'}{Q \trans{\out{h}{v}{}} Q'}{P| Q \trans{\tau} P'|Q'}
    \and
    \prftree[r]{$\fnames Q\cap\bnames\mu=\emptyset$}{P\trans\mu
      P'}{P|Q\trans\mu P'|Q}
    \\
    \prftree{\eval ev}{\eval fv}
    { P \trans{\mu} P'}{\ite{e=f}{P}{Q} \trans{\mu} P'}
    \and
    \prftree{\eval ev}{\eval f{v'}}{v\neq v'}
    { Q \trans{\mu} Q'}{\ite{e=f}{P}{Q} \trans{\mu} Q'}
  \end{mathpar}
  \caption{Labelled transition semantics.
    \\
    Symmetric versions of the
    rules for parallel composition and for sum have been omitted.}
  \label{f:lts}
\end{figure}

\section{Additional Material for Section~\ref{s:types}}
\label{sec:techlm}

\subsection{Properties of the type system}
\label{sec:typlm}

\begin{lm}
  \label{lm:fresh}
Suppose $\typ{\vfpl}{P}$
  and $a\notin \fnames P$.  Then also 
  $\typ{\resmin{\vfpl}{a}}{P}$.
\end{lm}
\begin{lm}[Invariance of typing w.r.t.\ structural congruence]
  \label{lm:typ_struct}
If  $\typ{\typpl{\vfpl}{\stack}}{P}$ and
  $P\equiv P'$, then also  $\typ{\typpl{\vfpl}{\stack}}{P'}$.
\end{lm}

\subsection{Properties of trace equivalence}
\label{sec:treqlm}

\begin{lm}[Shrinking]
  \label{l:shr}
Suppose
 $\vfopp[']\subseteq \vfopp$, $\vfpl$ and $\vfopp[']$ compatible and 
$P \trEqV \vfpl \vfopp Q$.
Then also 
$P \trEqV{\vfpl}{\vfopp'} Q$.
\end{lm}

\begin{df}[Projection] 
  Let $\vfopp$ be a visibility function and $N$ a set of names. We
  define a visibility function as follows:
  \[ \mathtt{proj}(N,\vfopp) :
    \begin{array}[h]{rcl}
      \Nho \cup \{\cur\} & \to & \mathcal{P}(\Nho)\\
      a\in \operatorname{dom}(\vfopp)\cap N & \mapsto & \vfopp(a) \cap N
    \end{array}\]
\end{df}
\begin{lm}[Extension]
  \label{l:tr_ext}
   Consider $\vfpl,\vfopp,\vfopp['],P,Q$ such that
   $\mathtt{proj}(\fnames{\vfopp,P,Q},\vfopp[']) = \vfopp$ and $\vfpl$ and $\vfopp[']$ are compatible. If
   $P \trEqV{\vfpl}{\vfopp} Q$ then
   $P \trEqV{\vfpl}{\vfopp[']} Q$. 
 \end{lm}

In the statement above, $\fnames{\vfopp,P,Q}$ stands for the union of
all names mentioned in $\vfopp$, together with the names occurring free
in $P$ and $Q$.

\subsection{Pruning Processes using Types}
\label{a:deadcode}

Consider a visibility function $\vfpl$ and
$S \subseteq \Nho$. 
We write \meets\alpha S when $\alpha$ is an input of the form $a(x,b)$
and either $a\in S$ or ($a\in\dom\vfpl$ and $\vfpl(a)\cap
S\neq\emptyset$).

We define
inductively the pruning function on processes that removes
all instances of the names in $S$ as follows: 
\begin{mathpar}
  \prune{S}{\vfpl}{\alpha. P} =
  \begin{cases}
    \nil \text{ if } \meets\alpha S\\
    \alpha. \prune{S}{\vfpl}{P} \text{ otherwise }
  \end{cases} 

  \prune{S}{\vfpl}{!\alpha. P} =
  \begin{cases}
    \nil \text{ if } \meets\alpha S\\
    !\alpha. \prune{S}{\vfpl}{P} \text{ otherwise }
  \end{cases}
  
  \prune{S}{\vfpl}{\out{\alpha}{}{}.P} =
  \out{\alpha}{}{}. \prune{S}{\vfpl}{P}
  
  \prune{S}{\vfpl}{\new a\ P} = \new a\  \prune{S}{\vfpl}{P}

  \prune{S}{\vfpl}{\new h\ P} = \new h\
  \prune{S}{\vfpl}{P}
  
  \prune{S}{\vfpl}{P|Q} = \prune{S}{\vfpl}{P} | \prune{S}{\vfpl}{Q}
  
  \prune{S}{\vfpl}{P+Q} =
  \prune{S}{\vfpl}{P} +\prune{S}{\vfpl}{Q}
  
  \prune{S}{\vfpl}{\nil} = \nil

  \prune{S}{\vfpl}{\omega} =
  \omega
  
  \prune{S}{\vfpl}{\ifte efPQ} =
    \ifte{e}{f}{\prune{S}{\vfpl}{P}}{\prune{S}{\vfpl}{Q}}

\end{mathpar}

\begin{lm}[Pruning]
  \label{lm:garbage_coll}
  Let $\vfopp, \vfpl,P$ such $\typ{\vfpl}{P}$ and $\vfopp$ and $\vfpl$
  are compatible and $\vfpl\cup \vfopp$ is transitive. Then for all $S\subseteq \Nho$ such that $S\cap
  (\vfopp\cup \vfpl (\cur)) = \emptyset$,
  $\typ{\resmin{\vfpl}{S}}{\prune{S}{\vfpl}P}$ and $P \trEqV{\vfpl}{\vfopp}
  \prune{S}{\vfpl}P$. 
\end{lm}
We actually remove more than the names in $S$. Indeed,  to
preserve sequentiality, we cannot  simply remove an output;   
we actually  remove
all inputs that may cause an output on the names in $S$.

The idea of the proof is to show that $S\cap (\vfopp \cup \vfpl
(\cur)) = \emptyset$ is an invariant throughout the trace; for this 
we use repeatedly
 transitivity of $\vfopp \cup \vfpl$. 
\begin{proof}
  Let $\mu_1\dots\mu_n$ a trace of $P$ under $\vfopp$, let $P_i,\vfopp_i$ be the processes and visibility functions given that trace and $\vfpl_i$ the visibility functions given by subject reduction.
  We show by induction on $i$ that there exists $\vfopp_i',\vfpl_i'$ such that $\vfopp_i'\supseteq \vfopp_i$ and $\vfpl_i' \supseteq \vfpl_i$ and
  $S\cap (\vfopp_i' \cup \vfpl_i')(\cur)= \emptyset$ and that $\vfpl_i'$ and $\vfopp_i'$ are compatible, have a transitive union and coincide on the intersection of their domains. The case $i=0$ is immediate, let $0\leq i <n$, we assume the induction hypothesis on $i$ and show it for $i+1$ by cases on $\mu_i$ :
  \begin{itemize}
  \item if $\mu_i = \tau$, or $\mu_i = \inpACT{h}{v}$ or $\mu_i = \out{h}{v}{}$ then $\vfopp_{i+1} = \vfopp_i$ and $\vfpl_{i+1} = \vfpl_i$, these cases are immediate by setting $\vfopp_{i+1}'= \vfopp_i'$ and $\vfpl_{i+1}'= \vfpl_i'$.
  \item if $\mu_i = \out{a}{v}{b}$ then $\vfopp_{i+1} = \vfopp_i\cons\cur \arr \vfopp_i(a),b$ and $\vfpl_{i+1} = \resmin{\vfpl_i}{\cur}\cons b\arr \vfpl_i(\cur),b$. 
    Let $\vfopp_{i+1}'= \vfopp_i'\cons\cur \arr \vfpl_i'(\cur),b \supseteq \vfopp_{i+1}$ because $a\in \vfpl_i(\cur)$ therefore $\vfopp_i(a)\subseteq \vfpl_i(\cur)$. And let $\vfpl_{i+1}'=\resmin{\vfpl_i'}{\cur}\cons b\arr \vfpl_i'(\cur),b$.\\
    $\vfopp_{i+1}'$    and $\vfpl_{i+1}'$ are compatible, their union is transitive and they coincide on the intersection of their domains.\\
    $(\vfopp_{i+1}' \cup \vfpl_{i+1}')(\cur) = (\vfopp_{i}' \cup \vfpl_{i}')(\cur),b$ where $b$ is fresh therefore $b\notin S$ and by induction hypothesis $S \cap (\vfopp_{i+1}' \cup \vfpl_{i+1}')(\cur)=\emptyset$.
\item if $\mu_i = a\langle v\rangle(b)$ the case is similar to the
  previous one, swapping $\vfopp$ and $\vfpl$. 
\end{itemize}
This ensures us that none of the $\mu_i$ are inputs or outputs on higher-order names in $S$. Moreover if $a\in \dom{\vfpl}$ and $\vfpl(a)\cap S\neq \emptyset$ then for all $i$, $a\notin (\vfopp_i \cup \vfpl_i)(\cur)$ otherwise we would have $\emptyset \neq \vfpl(a) \cap S = \vfpl_i(a) \cap S \subseteq (\vfopp_i \cup \vfpl_i)(\cur) \cap S = \emptyset$.\\
Therefore the trace $\mu_1\dots\mu_n$ is also a trace of $\prune{S}{\vfpl}{P}$ under $\vfopp$.
\end{proof}

\section{Section~\ref{s:labels}: Labelled Bisimilarity}\label{sec:lab_bisim}

\renewcommand{\typopp}[2]{ \textcolor{red}{#1}  }
\renewcommand{\bisim}[3]{ #2 \mathrel{\approx^{#1}} #3}
\renewcommand{\typtrans}[5]{\typPAIR{#1}{#2} \trans{#3}\typPAIR{#4}{#5}}
\renewcommand{\typpl}[2]{
 \textcolor{green}{#1}
}

  A  \emph{typed relation} is a set of triples $\underoppD{{
    \vfopp}}PQ$ intended to represent the visibility of the observer
for processes $P$ and $Q$ in the bisimulation game.


\begin{df}[Typed Bisimulation]
  A  typed relation $\mathcal{R}$ is a \emph{typed (weak) bisimulation} if
whenever $\underoppD{\vfopp}PQ \in \R$ : 
\begin{enumerate}
\item
if   $\typtrans{\typopp{\vfopp}{ }}{ P }{\mu}{ \typopp{\vfopp[']}{}}{
    P'}$ then there is $Q'$ such that
  $\typTrans{\typopp{\vfopp}{}}{ Q }{\mu}{ \typopp{\vfopp[']}{}}{ Q'}$
  and 
 $\underoppD{\vfopp'}{P'}{Q'} \in \R$;

\item  the converse, on the actions from $Q$.
\end{enumerate} 
\emph{Typed bisimilarity}, written $\approx$, is the largest typed
bisimulation. 
  For two $\vfpl$-processes $P,Q$ we write $P \bisEqV{\vfpl}{\vfopp} Q$  when
  $\underoppD{{\vfopp}{}}PQ \in {\approx}$ and $\vfpl$ and $\vfopp$ are compatible.
  In this case, we say that $P$ and $Q$ are \emph{bisimilar under $\vfopp$}.
\end{df}

\begin{ex}
  As for barbed equivalence the visibility of opponent also plays an
  important role in determining whether two process are bisimilar under
  some $\vfopp$. Consider 
$\vfpl$ and $\vfopp$ compatible, and 
a $\vfpl$-process  
  of the form $Q =
  P | a.\out{b}{}{}.R$ with $b\notin \mathtt{fn}(P)$ and $b \in
  \dom{\vfopp}$: 
  \begin{itemize}
  \item if $a \in \mathtt{fn}(P)$, we may have,  for example, $P=
    \out{a}{}{}.R'$, in which case $Q \trans{\tau} R' |
    \out{b}{}{}.R$; such derivative is not bisimilar with $P$ under 
    $\vfopp$, as only $R' | \out{b}{}{}.R$ can make an output on $b$;
  \item if $a\in \vfopp(\cur)$, then 
 we have $Q \trans{a} P| \outC{b}.R$, which is not bisimilar to any
 $P'$ such that $P \trans{a}P'$,  for the same reason as in the previous case;
  \item however, if both $a\notin \mathtt{fn}(P)$ and $a\notin
    \vfopp(\cur)$, then  $Q$ cannot
    make an input on $a$,  and therefore we have $Q\bisEqV{\vfpl}{\vfopp} P$.
  \end{itemize}
\end{ex}

The analogues of Lemmas~\ref{lm:res}-\ref{lm:parcomp_tr} and~\ref{lm:garbage_coll} also hold for 
typed bisimilarity. 
\begin{lm}[Restriction]
  \label{lm:res_bis}
 If 
$P \bisEqV \vfpl \vfopp Q$,
then also 
$\res a P \bisEqV{\resmin{\vfpl}{a}}{ \resmin{\vfopp}{{a}}} \res a Q$.
\end{lm}
\begin{lm}[Input]
  If $P \bisEqV \vfpl \vfopp Q$ and
$\vfopp \NOTinDom \cur$, 
then also $a(x,b) .P \bisEqV{\resmin{\vfpl}{\cur,b}}{ (\resmin{\vfopp\cons \cur \arr \vfopp(b))}{b}} a(x,b).Q$.
\end{lm}
\begin{lm}[Output]
  If $P \bisEqV\vfpl \vfopp Q$ and
$\vfopp \NOTinDom \cur$, 
then also $\out{a}{e}{b}. P \bisEqV{\resmin{(\vfpl\cons \cur \arr \vfpl(b))}{b}}{ \resmin{\vfopp}{\cur,b}} \out{a}{e}{b}.Q$.
\end{lm}

\begin{lm}[Parallel composition]
  \label{lm:parcomp_bis}
Suppose $P \bisEqV \vfpl \vfopp Q$
and 
  $(\vfopp[_1], \vfopp[_2])\in \mathtt{split}(\vfopp)$. 
Then 
  $\typ{\vfopp[_2]}{R}$
implies
 $P| R \bisEqV{(\vfpl\cup \vfopp_2)^+}{\vfopp_1} Q|R$.
\end{lm}

\begin{lm}[Pruning]
  \label{lm:garbage_coll_bis}
  Let $\vfopp, \vfpl,P$ such $\typ{\vfpl}{P}$ and $\vfopp$ and $\vfpl$
  are compatible. Then for all $S\subseteq \Nho$ such that $S\cap
  (\vfopp\cup \vfpl (\cur)) = \emptyset$,
  $\typ{\resmin{\vfpl}{S}}{\prune{S}{\vfpl}P}$ and $P \bisEqV{\vfpl}{\vfopp}
  \prune{S}{\vfpl}P$. 
\end{lm}

Technical lemmas similar to those for trace equivalence also holds,
such as the following ones.

\begin{lm}[Shrinking]
  \label{l:shr:bis}
Suppose
 $\vfopp[']\subseteq \vfopp$, $\vfpl$ and $\vfopp[']$ compatible and
$P \bisEqV\vfpl \vfopp Q$.
Then also 
$P \bisEqV{\vfpl} {\vfopp'} Q$.
\end{lm}

\begin{lm}[extension]
   Consider $\vfopp,\vfopp['],P,Q$ such that
   $\mathtt{proj}(\fnames{\vfopp,P,Q},\vfopp[']) = \vfopp$ and $\vfpl$ and $\vfopp'$ are compatible. If
   $P \bisEqV{\vfpl}{\vfopp} Q$ then
   $P \bisEqV{\vfpl}{\vfopp[']} Q$. 
 \end{lm}

We report the characterisation results.
Soundness follows from the substitutivity properties of labelled
bisimilarity. For completeness, as usual with characterisations of 
barbed equivalence, the proof  uses induction, via 
an  inductive stratification of
bisimilarity, and requires image-finiteness of processes 
(that is, up-to $\alpha $-conversion, the set of possible
type-allowed  weak transitions emanating from a process is finite). 
The image-finite condition is needed 
in the 
 characterisation of bisimilarity via the inductive 
stratification.
Also, and again in line with  proofs about barbed relations, 
   the  proof exploits the  presence of the sum operator and of
success signals that can be emitted by a testing context whenever
needed. 

\begin{thm}[Soundness of  bisimilarity  for barbed equivalence]
 Suppose $\typ{\vfpl}{P,Q}$
with 
$P\bisEqV{\vfpl}{\vfopp} Q$.
Then also 
 $P
 \wbe \vfpl \vfopp
 Q$.
\end{thm}

\begin{thm}[Completeness of bisimilarity  for barbed equivalence]
If  $P$ and $Q$ are image-finite processes, and if
 $P
 \wbe  \vfpl \vfopp
 Q$,
 then 
$P\bisEqV{\vfpl}{\vfopp} Q$.
\end{thm}

\renewcommand{\typpl}[2]{\langle
 \textcolor{green}{#1}
 |\textcolor{green} {#2} \rangle
}
\renewcommand{\typopp}[2]{[
  \textcolor{red}{#1} |
  \textcolor{red}{#2}] 
}
\renewcommand{\typtrans}[5]{{#1}{#2} \xrightarrow{#3}
  {#4}{#5}}
\renewcommand{\typTRANS}[5]{{#1}{#2} \xRightarrow{{#3}}{#4}{#5}}
\renewcommand{\typTrans}[5]{{#1}{#2} \xRightarrow{\hat{#3}}{#4}{#5}}

\section{Additional Material for Section~\ref{s:wb}}\label{sec:bis_wb}

\subsection{About
  Example~\ref{ex:INC}}\label{a:reentrant:call}

The process discussed in Example~\ref{ex:INC} can be seen as the
$\pi$-calculus counterpart of the following ML program, which might be
more readable for a reader familiar with functional languages:
\vskip .2cm 

\noindent\texttt{let h = ref 0 in \\
  let b () = h := !h +1 in\\
  begin\\
\mbox{}~\quad    a b;\\
\mbox{}~\quad    let x = !h in\\
\mbox{}~\quad      c x;\\
\mbox{}~\quad      let y = !h in\\
\mbox{}~\quad      if x=y then loop() else ()\\
  end
}

\paragraph*{Re-entrant call.}

The following example first shows an equality similar to that of
Example~\ref{ex:INC}; and then develops the example so to show 
 how re-entrant calls may sometimes 
break `expected'    visibility boundaries.

We define two processes $P_1$ and $P_2$, which access a reference $h$
(implemented using a first-order name), and can call a server
$b$. When calling $b$, the processes 
 first pass the name $\mathtt{incr}$ of a
server that increments the value stored in $h$. 
Once the
call on $b$ is answered the processes read the value in $h$ and make another
call to $b$ passing the identity function. After that,
the processes read again the value of $h$. Finally, $P_1$ outputs the first value
read in $h$ while $P_2$ outputs the second one. 

\begin{align*}
  P_i \defi \out{b}{}{\mathtt{incr},q_1}. & q_1. h(x_1). (\out{h}{x_1}{} \\
                                    & | \ \out{b}{}{\mathtt{id},q_2}.  \\
                                    & \hspace{1em} q_2. h(x_2). (\out{h}{x_2}{} \\
                                    & \hspace{2em} | \ \out{p}{x_i}{}))\\
\end{align*}
We first compare the processes
$$\new h( \out{h}{0}{} |\ a(b,p).P_1 |\
Q_{\mathtt{incr}} |\ Q_{\mathtt{id}})
\quad\text{and}\quad
\new h( \out{h}{0}{}|\ a(b,p).P_2 
|\ Q_{\mathtt{incr}} |\ Q_{\mathtt{id}}),$$
\noindent
where $Q_{\mathtt{incr}}$
(resp.\ $Q_{\mathtt{id}}$) is the encoding of function \texttt{incr}
(resp.\ \texttt{id}).

Here, the only way to change the
value stored in $h$ is by calling $\mathtt{incr}$, which is not
visible by the observer after the input on $q_1$. These two
processes are bisimilar under any visibility.

Then we compare the processes
$$Q_1 \defi \new h( \out{h}{0}{} |\ !a(b,p).P_1
|\ Q_{\mathtt{incr}} |\ Q_{\mathtt{id}}))
\quad\text{and}\quad
Q_2 \defi \new  h( \out{h}{0}{}|\ !a(b,p).P_2
|\ Q_{\mathtt{incr}} |\ Q_{\mathtt{id}}).$$
We are be able to distinguish them using a re-entrant call :
\begin{align*}
  R \defi \out{a}{}{b,p}. (  & b(\mathtt{incr},q).\out{a}{}{d,r}.( !d(-,s).\out{\mathtt{incr}}{}{t}.t.\out{s}{}{} \\
                              & \hspace{8em} |\  r(n). \ite{n=1}{\out{q}{}{}}{\omega}) \\
                         &|\  p(m) )
\end{align*}
The idea is to call $a$ a second time when $\mathtt{incr}$ is still in
the observer's visibility; this means that the new server passed to $a$ has
access to $\mathtt{incr}$ during both calls. With the observer $R$,
only $Q_2$ will be able to send a success signal. Therefore, for any
visibility function $\vfopp$ such that $a\in \vfopp(\cur)$, $Q_1$ and
$Q_2$ are not bisimilar under visibility $\vfopp$. 

The processes $Q_1$ and $Q_2$ can be seen as the $\pi$-calculus
counterparts of the following ML program. 
(This  example, with re-entrant calls, has been inspired by similar
examples in game semantics of functional languages):

\noindent\texttt{let h = ref 0 in \\
  let incr () = h := !h +1 in\\
  let id x = x in\\
  let a b =\\
\mbox{}~\quad    b incr;\\
\mbox{}~\quad    let x$_1$ = !h in\\
\mbox{}~\quad      b id;\\
\mbox{}~\quad      let x$_2$ = !h in\\
\mbox{}~\quad      x$_i$
}

These programs can be distinguished by the following ML program that corresponds to the process $R$ :

\noindent\texttt{
  let b f = \\
  \mbox{}~\quad let d \_ = f () in\\
  \mbox{}~\quad let n = a d in \\
  \mbox{}~\quad if n = 1 then () else $\omega$ \\
  in\\
  a b
  }

The idea is the same as before: \texttt{d} calls \texttt{f} which will
be \texttt{incr} in the first call to \texttt{b}; therefore at each
call to \texttt{d} the reference is incremented. As a consequence, 
 first program returns 1 while the second one returns 2.

\subsection{Typed bisimilarity}

\begin{df}[Typed Bisimulation]
  A  typed relation $\mathcal{R}$ is a \emph{typed (weak) bisimulation} if
whenever $\underoppD{\typopp{\vfopp}{\kcats}}PQ \in \R$ : 
\begin{enumerate}
\item
if   $\typtrans{\typopp{\vfopp}{ \kcats}}{ P }{\mu}{ \typopp{\vfopp[']}{\kcats[']}}{
    P'}$ then there is $Q'$ such that
  $\typTrans{\typopp{\vfopp}{\kcats}}{ Q }{\mu}{ \typopp{\vfopp[']}{\kcats[']}}{ Q'}$
  and 
 $\underoppD{\typopp{\vfopp'}{\kcats'}}{P'}{Q'} \in \R$;

\item  the converse, on the actions from $Q$.
\end{enumerate} 
\emph{Typed bisimilarity}, written $\approx$, is the largest typed
bisimulation. 
  For two $\typpl{\vfpl}{\stack}$-processes $P,Q$ we write $P \bisEqV{\typpl{\vfpl}{\stack}}{\typopp{\vfopp}{\kcats}} Q$ when
  $\underoppD{\typopp{\vfopp}{\kcats}}PQ \in {\approx}$ and $\typpl{\vfpl}{\stack}$ and $\typopp{\vfopp}{\kcats}$ are compatible.
\end{df}

\begin{thm}[Soundness of  bisimilarity  for barbed equivalence]
 Suppose $\typ{\typpl{\vfpl}{\stack}}{P,Q}$
with 
$P\bisEqV{\typpl{\vfpl}{\stack}}{\typopp{\vfopp}{\kcats}} Q$.
Then also 
 $P
 \wbe{\typpl{\vfpl}{\stack}}{\typopp{\vfopp}{\kcats}}
 Q$.
\end{thm}

\begin{thm}[Completeness of bisimilarity  for barbed equivalence]
If  $P$ and $Q$ are image-finite processes, and if
 $P
 \wbe{\typpl{\vfpl}{\stack}}{\typopp{\vfopp}{\kcats}}
 Q$,
then 
$P\bisEqV{\typpl{\vfpl}{\stack}}{\typopp{\vfopp}{\kcats}} Q$.
\end{thm}

\subsection{Properties of traces: views and well-bracketing}

Along the lines of the material presented in Section~\ref{s:view}, we
can reason about the traces produced by typable processes, based on
the notion of justification.
%
For this, we adapt Definition~\ref{d:view}  to define views for traces.
%
The notion of \emph{respecting views}
(Definition~\ref{d:respect:view}) is also adapted straightforwardly,
which allows us to obtain the analogue of Lemma~\ref{l:view}.
\begin{lm}
  Consider $\vfpl,\vfopp,\stack,\kcats,P$ such that $\typ{\typpl{\vfpl}{\stack}}{P}$ and $\typopp{\vfopp}{\kcats}$ is
  compatible with $\typpl{\vfpl}{\stack}$. Let $\Vset=(\vfopp \cup \vfpl)(\cur)$,
  then every
  trace of $P$ under $\typopp{\vfopp}{\kcats}$ respects views with starting
  visibility \Vset.
\end{lm}

As highlighted in~\cite{Hirschkoff_2021}, 
continuation names allow us to formulate a property of
well-bracketing for traces, enforced by our type system. 
In contrast with~\cite{Hirschkoff_2021}, however, we 
 we do not
impose input receptiveness on continuation names.

\begin{df}[Well-bracketing]
  \label{d:wb}
An action of the form $\inpACT a{v}(b,p)$ or $\out a{v}{b,p}$ is
called a \emph{question}.
An action of the form $\inpACT p{v}(b)$ or $\out p{v}{b}$ is
called an \emph{answer}.

A trace $\mu_1\dots\mu_n$ is \emph{well-bracketed} if for all
$i < j$, if $\mu_i$ is a question and $\mu_j$ is an answer with
$\mu_i\njustif \mu_k$ and $\mu_k \njustif \mu_j$ for all $i < k < j$,
then $\mu_i \justif \mu_j$.
\end{df}



A stack $\stack$ is  \emph{clean} if each name has at most one
occurrence, with any tag, in $\stack$.
 A trace $\mu_0\ldots\mu_n$ is a \emph{typed trace emanating
  from $P_0$} 
  if there are 
	$ \vfpl[_1], \sigma_1,\dots,\vfpl[_n],\sigma_n, P_1,\dots,P_n$
	such that for all  $0\leq j < n$ we are in one of the cases
        given by Theorem~\ref{t:SR:wb}, which gives
 ${\typpl{\vfpl[_{j+1}]}{\stack_{j+1}}}$ when
  $P_j\xrightarrow{\mu_{j+1}}P_{j+1}$.

\begin{lm}
  Suppose 
  $\typ{\typpl{\vfpl_0}{\stack_0}}{P_0}$, where $\stack_0$ is clean.
  All typed traces emanating from $P_0$ are well-bracketed.
\end{lm}

\finish{above: all *typed* traces? we would need type-allowed traces for sequentiality but we did not define it // suggestion : traces under $\vfopp$ forall $\vfopp$}

 \end{document}

\newpage \input{todo}

\end{document}



\input{related}
\input{cut} 
\end{document}